\documentclass[12pt]{iopart}

\pdfoutput=1
\usepackage[utf8]{inputenc}
\usepackage[USenglish]{babel}

\usepackage{fancyhdr}

\usepackage{amssymb,amsthm,amsfonts}%amsmath
\usepackage{MnSymbol}
\usepackage{mathrsfs}
\usepackage{slashed}
\usepackage{graphicx}
\usepackage{subfigure}
\usepackage{color,rotating}
\usepackage{dsfont}
\usepackage{setspace}
\usepackage{physics}
\usepackage{verbatim}
\usepackage{hyperref,doi}
\usepackage[export]{adjustbox}
\usepackage[normalem]{ulem}
\usepackage{bbold}

\usepackage[labelfont=bf, font=footnotesize]{caption}

\makeatletter
\def\l@subsubsection#1#2{}
\makeatother

\newcommand{\beq}{\begin{equation}}
\newcommand{\eeq}{\end{equation}}
\newcommand{\beqa}{\begin{eqnarray}}
\newcommand{\eeqa}{\end{eqnarray}}
\newcommand{\bfc}{\begin{figure}[t]\begin{center}}
\newcommand{\efc}{\end{center}\end{figure}}

\def\fig#1{Fig.~\ref{#1}}

\def\sec#1{Section~\ref{#1}}
\def\0#1#2{\frac{#1}{#2}}  %% fractions

\graphicspath{{./figures/}}
\newcommand{\be}{\begin{eqnarray}}
\newcommand{\ee}{\end{eqnarray}}

\newcommand{\del}{\partial}

\begin{document}

\title[Unsupervised Learning of Rydberg Atom Array Phase Diagram with Siamese Neural Networks]{Unsupervised Learning of Rydberg Atom Array Phase Diagram with Siamese Neural Networks}

\author{Zakaria Patel}
\address{Department of Engineering Physics, McMaster University, Hamilton, Ontario L8S 4L8, Canada}
\author{Ejaaz Merali}
\address{Department of Physics and Astronomy, University of Waterloo, Waterloo, Ontario N2L 3G1, Canada \\ Perimeter Institute for Theoretical Physics, Waterloo, Ontario N2L 2Y5, Canada}
\author{Sebastian J. Wetzel}
\address{Perimeter Institute for Theoretical Physics, Waterloo, Ontario N2L 2Y5, Canada}

%%%%%%%%%%%%%%%%%%%%%%%%%%%%%%%%

\begin{abstract}
We introduce an unsupervised machine learning method based on Siamese Neural Networks (SNN) to detect phase boundaries. This method is applied to Monte-Carlo simulations of Ising-type systems and Rydberg atom arrays. In both cases the SNN reveals phase boundaries consistent with prior research. The combination of leveraging the power of feed-forward neural networks, unsupervised learning and the ability to learn about multiple phases without knowing about their existence provides a powerful method to explore new and unknown phases of matter.
\end{abstract}
\vspace{2pc}
\maketitle

%%%%%%%%%%%%%%%%%%%%%%%%%%%%%%%% 
\section{Introduction}

Machine learning (ML) algorithms enable computers to learn from experience and generalize their gained knowledge to previously unknown settings. It is perhaps the most transformative technology of the early 21th century. The ability to recognize objects in images \cite{krizhevsky2012imagenet} or translate languages \cite{goldberg2016primer} without being explicitly programmed for this task, highlights the enormous potential of machine learning.

In recent years the physical sciences have adopted machine learning based algorithms to explore complex questions. Many methods have been designed to solve problems beyond the scope of data science, and have now the potential to revolutionize physics. The most prominent examples of promising tasks that have been tackled include finding phase transitions \cite{Carrasquilla2017,Nieuwenburg2017,Wang2016,Wetzel2017,Zhang2017,Schindler2017,Hu2017,Ohtsuki2017,Broecker2017,Deng2017,Chng2017}, reconstructing or simulating quantum systems \cite{Torlai2016,Carleo2017,Inack2018,Hibat-Allah2020,Carrasquilla2019,Ferrari2019,Sharir2020} and rediscovering physical concepts \cite{Schmidt2009,Iten2020,Wetzel2017a,Ponte2017,wetzel2020discovering,Greitemann2019,Mototake2019,Udrescu2019,krenn2017entanglement,krenn2022scientific}. 
All these advances are summarized in review articles directed at different audiences. A didactical review to the most modern techniques can be found in \cite{dawid2022modern}, a review article focused on the applications of machine learning to examine quantum matter \cite{carrasquilla2020machine} and a broad overview across different physical disciplines is summarized in \cite{carleo2019machine}.

The current manuscript contributes to the development of methods that automatize the calculation of phase diagrams with little to no human prior knowledge about the nature of the underlying phases.
The subfield of automated phase recognition can be subdivided into two categories: supervised and unsupervised phase recognition.

In the first case, the operating scientists are aware of the of the possible phases and have a rough estimate of where these phases are positioned in the phase diagram; however, they are unsure about the exact location of the phases and the transitions between them. Supervised learning of phase transitions can be based on different machine learning algorithms. It was initially introduced using convolutional neural networks \cite{Carrasquilla2017}, which are to this day the most powerful and robust tools to learn accurate physical phase boundaries. There are hybrid methods that build upon purposely mislabelling phase classes \cite{Nieuwenburg2017}, methods that are built upon support vector machines \cite{Ponte2017}, and other powerful frameworks \cite{Zhang2017,huembeli2018identifying,arnold2022replacing,van2018learning,hsu2018machine,vargas2018extrapolating,bachtis2020extending}. These methods have demonstrated success across a wide range of physical systems, from simple spin lattices, over strongly correlated quantum systems, up to lattice gauge theories \cite{Schindler2017,Ohtsuki2017,Broecker2017,Chng2017,Wetzel2017a,beach2018machine,zhang2019machine,Suchsland2018,kim2018smallest,lian2019machine,dong2019machine,giannetti2019machine,ohtsuki2020drawing,casert2019interpretable,zhang2019few,singh2019application}. 

The second category contains unsupervised phase recognition algorithms. These algorithms are useful when the researcher who is employing these tools is unaware of the underlying phase structure, meaning they do not know about the existence or location of certain phases and thus cannot supply this information to the machine learning algorithm. The simplest unsupervised phase recognition scheme is based on principal component analysis \cite{Wang2016} and the most widely used unsupervised scheme that leverages the power of artificial neural networks is based on variational autoencoders \cite{Wetzel2017}. These methods have been examined, enhanced \cite{Hu2017,arnold2021interpretable,liu2018discriminative,huembeli2019automated} and successfully applied to many systems in physics and materials science \cite{kaming2021unsupervised,alexandrou2020critical,yin2021neural,Greitemann2019,greplova2020unsupervised,kharkov2020revealing,ch2018unsupervised,wang2017machine,kottmann2020unsupervised,jadrich2018unsupervised,che2020topological}. Compared to supervised algorithms, unsupervised methods usually have the drawback of lacking accuracy in determining phase boundaries \cite{Wetzel2017}, or restricting the kinds of order parameters that can be learned \cite{Wang2016}.

We introduce an unsupervised machine learning method to discover phase transitions based on Siamese neural networks (SNN). Siamese networks were initially introduced for fingerprint and signature identification \cite{bromley1993signature,baldi1993neural}. Instead of predicting a certain class, Siamese networks predict if two inputs belong to the same class. Hence, these networks can be used for multi- or infinite class classification by comparison to anchor data points whose label is known. Although Siamese networks are very powerful, they have experienced little use in the physical sciences. So far Siamese networks have been employed to discover symmetry invariants and conserved quantities \cite{wetzel2020discovering}. While Siamese neural networks are supervised machine learning algorithms, our proposed phase recognition method is unsupervised, in the sense that it does not require any phase information. This apparent contradiction is reconciled in \sec{sec:SNNphases}.

While we initially present our phase recognition method using the example of two stacked Ising models exhibiting four different phases, we demonstrate the power of this method by examining the phase diagram of the Rydberg atom array.
Rydberg atom arrays are a powerful platform for experimental realizations of quantum many-body phenomena \cite{Henriet2020quantumcomputing,Browaeys_2020,kalinowski2021bulk}. Neutral atoms are typically arranged via optical tweezers to construct various physical lattices at varying interaction strengths which give rise to rich phase diagrams \cite{ebadi2021}.
Such systems have already been examined with the help of machine learning algorithms. The phase diagram has been revealed by a combined effort of unsupervised and supervised methods \cite{miles2021machine}. Experimental states have been reconstructed using neural network based tomography \cite{torlai2019integrating}. Ground states have been calculated \cite{carrasquilla2021use} and simulated measurement data has been used for pre-training variational wave functions \cite{czischek2022data}.

The paper is structured in the following order: we first introduce the models we are examining with our new Siamese network based framework. These models include a stacked Ising model and the Rydberg atom array. Subsequently, we describe how we prepare the input data using Monte Carlo simulations. We describe how Siamese neural networks are constructed and trained, and develop our framework to do unsupervised learning of phase transitions. Then we apply our method to both models and present the results in the form of the predicted underlying phase structure. Finally, we summarize our findings and place our method into the broader context of machine learning tools for phase recognition.

\newpage

\section{Models}

\subsection{Stacked Ising Model}
The Ising model on the square lattice is a simple, well studied, and exactly solvable model from statistical physics that exhibits a phase transition. Thus, it provides the ideal starting point for benchmarking the performance of a phase recognition algorithm. Its Hamiltonian is
\begin{align}
H(S)=-J \sum_{<ij>_{nn}} s_i s_j + h \sum_i s_i,
\end{align}
a function of spin configurations $S=(s_1,...,s_n)$. In the following, we focus on the ferromagnetic Ising
model $J = 1$ and set the external field $h=0$.

SNN phase recognition is an algorithm that is able to detect multiple phases in an unsupervised manner. Hence, we trivially combine two Ising models by overlaying them on the same lattice where we can tune each temperature $T,\tilde{T}$, or in other words the ratios $J/k_BT$ and $\tilde{J}/k_B \tilde{T}$, independently. The combined Hamiltonian 
\begin{align}
H(S)=-J \sum_{<ij>_{nn}} s_i s_j -\tilde{J} \sum_{<ij>_{nn}} \tilde{s}_i \tilde{s}_j
\end{align}
acts on lattices containing two spins per site, $S=((s_1,\tilde{s_1}),...,(s_n,\tilde{s_n}))$. Since there is no interaction between them, phase transitions trivially occur at lines $(T,\tilde{T})\approx (\ \cdot \ ,2.269)$ and $(T,\tilde{T})\approx (2.269,\ \cdot \ )$ in the phase diagram.

\begin{figure}[htb!]
\begin{center}

\includegraphics[width=0.3\textwidth]{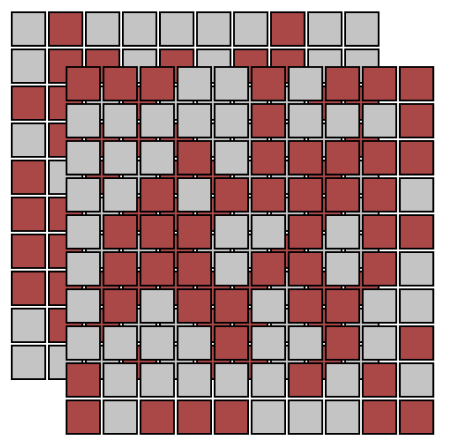}\\
\caption{Combining the Ising lattices involves the simple stacking shown in this figure. The red sites indicate spin-up, and the grey sites correspond to spin-down. The two lattice configurations are independently sampled at temperatures $T$ and $\tilde{T}$ to produce a single stacked configuration at $(T,\tilde{T})$.}
\label{fig:isingOverlay}
\end{center}
\end{figure}

\subsection{Rydberg Array}

A common Hamiltonian that can be implemented by Rydberg arrays has a form similar to that of a Transverse Field Ising Model, meaning it is sign-problem free and thus amenable to simulation on classical computers. The Hamiltonian acts on a collection of atoms which individually act like 2-level systems, having a ground-state $\ket{g} \equiv \ket{0}$ and an excited, so-called \textit{Rydberg} state $\ket{r} \equiv \ket{1}$. The atoms are subject to a long-range interaction which is described by a van der Waals (vdW) interaction of the form $V_{ij} \sim \abs{\mathbf{r}_i - \mathbf{r}_j}^{-6}$ that penalizes atoms that are simultaneously in the Rydberg state \cite{Browaeys_2016}. Additionally, the atoms are subject to coherent laser fields: a \textit{detuning} $\delta$ which acts like a chemical potential, driving atoms into their Rydberg states, and a \textit{Rabi oscillation} with frequency $\Omega$ which excites ground-state atoms and de-excites atoms in Rydberg states.
\begin{equation}\label{eq:rydberg_hamiltonian}
    H = \sum_{i<j} V_{ij}n_i n_j - \delta \sum_i n_i + \frac{\Omega}{2}\sum_i \sigma_i^x
\end{equation}
where $n_i = \ketbra{1}_i$ is the occupation/number operator, and $\sigma_i^x = \ketbra{1}{0}_i + \ketbra{0}{1}_i$. The interaction strength is typically parametrized in terms of a \textit{Rydberg blockade radius} $R_b$, which describes an effective radius within which two simultaneous Rydberg excitations are heavily penalized: $V_{ij} = \Omega (R_b / \abs{\mathbf{r}_i - \mathbf{r}_j})^6$ \cite{samajdar2020,merali2021stochastic}.

\section{Methods}
\subsection{Monte-Carlo Simulation}

The well-known single-spin-flip Metropolis algorithm is used to generate importance sampled Monte Carlo configurations of the Ising model on a square lattice of size $20 \times 20$ with periodic boundary conditions \cite{newmanb99}. After initializing a random lattice, we evolve the simulation for 7168 MC steps between drawing samples. It is important to note that neural networks can pick up on any residual correlations, thus relying on conventional auto-correlation measures to determine the independence of lattice configurations is not enough. We produce 92 independent configurations at each of 100 temperatures ranging from $T_{min} = 1.53$ to $T_{max} = 3.28$. This naturally translates to $92\times92 = 8464$ samples for the stacked Ising model at each temperature pair $(T,\tilde T)$.

For the Rydberg system, we make use of a recent Quantum Monte Carlo method \cite{merali2021stochastic} to generate occupation basis samples of the Rydberg Hamiltonian.
The QMC simulation is based on a power-iteration scheme which projects out the ground-state of the Hamiltonian:
\begin{equation}
    \ket{E_0} \approx (C - H)^M \ket{+}^{\otimes N},
\end{equation}
where $\ket{+}$ is the positive eigenstate of $\sigma_x$, $N$ is the number of lattice sites, $C$ is a constant energy shift used to cure the sign-problem emerging from the diagonal part of the Hamiltonian, and $M$ is called the projection length.
We perform our simulations on a $16\times 16$ square lattice with open boundaries at various parameter values. Unlike previous DMRG-based studies\cite{samajdar2020,ebadi2021} we do not impose a truncation on the vdW interaction. We take our projection length $M$ to be 100,000 which we found was more than enough to accurately converge to the ground-state over the parameter sets which were simulated. For our simulations we fixed $\Omega = 1$ and performed scans over $R_b \in \{1.0, 1.1, \ldots, 1.8\}$ and $\delta \in \{0.5, 0.6, \ldots, 2.9\}$. 

%\zak{We ended up doing $R_b \in \{1.0, 1.1, \ldots, 1.8\}$, and $\delta \in \{0.5, 0.6, \ldots, 2.9\}$}

To generate the occupation basis data for the SNNs, we first perform 100,000 Monte Carlo update steps to allow the chain to reach equilibrium. We then record one sample every 10,000 steps in order to eliminate any possible autocorrelation between successive samples. Each Monte Carlo step consists of a diagonal update step, followed by a cluster update step in which all possible line-clusters are built deterministically and flipped independently according to a Metropolis condition; see \cite{merali2021stochastic} for further details. Additionally, at each point in parameter space $(R_b, \delta)$ we run 3 independent Markov chains. The chains are allowed to evolve until each has generated 400 samples, giving a total of 1200 independent samples for each parameter pair.

\subsection{Siamese Neural Networks}
\begin{figure}[htb!]
\begin{center}
\includegraphics[width=1\textwidth]{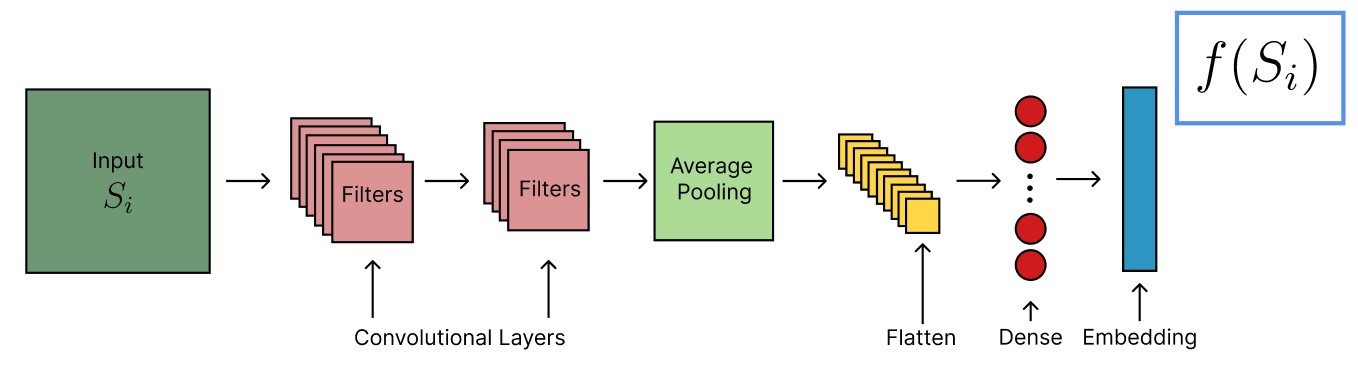}\\
\caption{The architecture of the convolutional subnetwork $f$ in the SNN $F$. The input $S_i$ is first passed through 2 convolutional layers, followed by an average pooling to capture a quantity characterized by an average. The result is flattened in order to feed it to a dense layer. This dense layer is then connected to an embedding layer.}
\label{fig:SNNarch}
\end{center}
\end{figure}

Artificial neural networks are directed graphs that have the ability to learn an approximation to any smooth function $f(x)=y$ given sufficiently many parameters. A neural network is built by successively applying matrix multiplications characterized by weights $w_{ij}^L$ that are offset by biases $b_i^L$. ($i,j$ are neuron indices in different layers $L$). Between subsequent matrix multiplications there is a non-linear activation function, common choices of which are sigmoid or rectified linear units. A neural network is trained by applying it to a data set and optimizing the network parameters to minimize a certain objective function using gradient descent.

Siamese neural networks (SNN) were introduced to solve an infinite class classification problem as it occurs in finger print recognition or signature verification \cite{bromley1993signature,baldi1993neural}. Instead of assigning a class label to a data instance, the SNN compares two data points and determines their similarity. A solution to the infinite class classification problem is obtained by calculating the predicted similarity between labelled anchor data points and a new unlabelled data instance.

\begin{figure}[htb!]
\begin{center}
\includegraphics[width=0.8\textwidth]{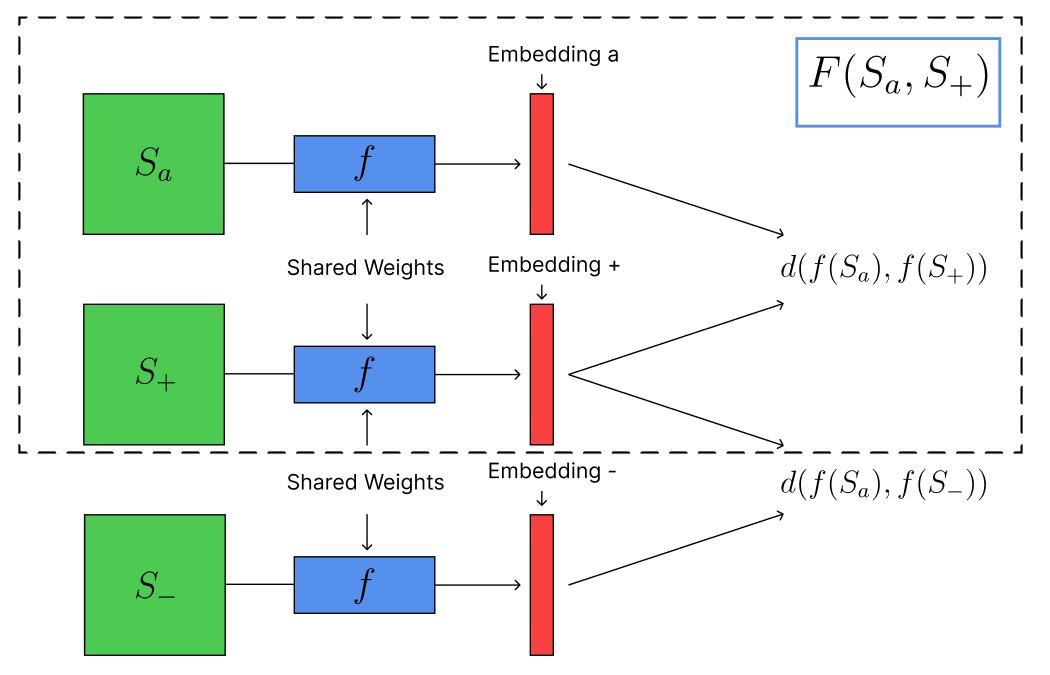}\\\caption{A Siamese neural network (SNN) $F$ takes a pair of input samples $(S_i,S_j)$ and predicts their similarity $d(f(S_i),f(S_j))$ The training architecture effectively mimics two SNNs acting on a triplet $(S_a,S_+,S_-)$ of samples. $S_a$ represents the anchor input, $S_+$ the positive sample from the same class as the anchor, and $S_-$ the negative sample from a different class. The shared convolutional subnetworks $f$ map their inputs into an embedding space. The SNN is trained to minimize distances $d(f(S_a), f(S_+))$, while maximizing distances $d(f(S_a), f(S_-))$. During the inference stage, we discard one branch of this network to obtain a pairwise comparison between the anchor and a new unknown sample.}
\label{fig:SNNDesign}
\end{center}
\end{figure}

A SNN $F(S_i,S_j)$ (\fig{fig:SNNDesign}) consists of two identical sub-networks $f$ (\fig{fig:SNNarch}) which project an input pair into a latent embedding space. The similarity of two inputs is determined based on the distance in embedding space.
\begin{align}
F(S_i,S_j)=d \left(   f(S_i) , f(S_j)   \right)
\end{align}
Possible distance metrics $d$ must be chosen according to the problem at hand, in our case we chose the average squared Euclidean distance on the unit sphere which is equivalent to the cosine distance.

\begin{equation}
d(f_1, f_2) = \frac{\sum_{i=0}^N (f_{1i}-f_{2i})^2}{N},
\end{equation}

Here $f_{1i},f_{2i}$ denotes different components in an $N$ dimensional embedding space. Instead of training the SNN on paired training data, an effective way to train Siamese neural networks is through minimizing contrastive loss functions involving triplets $(S_a,S_+,S_-)$ of data points

\begin{equation}
\mathcal{L}=\max(d(f(S_a), f(S_+)) - d(f(S_a), f(S_-)) + \alpha, 0),  \label{eq:contrastive_loss}
\end{equation}

where the hyperparameter $\alpha$ is chosen such that the neural network is prevented from learning trivial embeddings. The intuition behind $\alpha$ is its interpretation as a margin to encourage separation between the anchor and the negative embedding in the embedding space \cite{tripletLoss}. Minimizing $\mathcal{L}$ requires minimizing the distance between the anchor and positive sample $d(f(S_a), f(S_+))$, while maximizing the distance between the anchor and negative sample $d(f(S_a), f(S_-)$. $\max ( \cdot,0 )$ prevents the neural network from learning runaway embeddings $f(S_i)_i\rightarrow \infty$.

\subsection{Model Architecture}
\label{sec.architecture}
The explicit model architecture for $f$ \fig{fig:SNNarch} depends on the underlying data set. In the case of the Ising model, $f$ consists of two 2-D convolutional layers, both with stride $(1, 1)$, a kernel size of $(3, 3)$, and with 6 and 10 filters, respectively. Each layer is fed into a ReLU activation function. The resulting image dimensions are $(16, 16)$. This is followed by a $(16, 16)$ average pooling layer. The output is flattened and fed to a dense layer with 10 neurons. Subsequently, we feed this output to the embedding layer, which also contains 10 neurons and a sigmoid activation function. The embedding is normalized to unit length under euclidean norm. 

The model architecture for examining the Rydberg system is similar to the Ising model. In this case, we begin with two 2-D convolutional layers, both with stride (1, 1), a kernel size of (3, 3), and with 6 and 4 filters, respectively. The resulting image dimensions are (12, 12). As before, we subsequently apply an average pooling layer, but this time of dimension (12, 12). The remainder of the architecture is identical to before. 

We use the Adam optimizer to train our neural network. Furthermore, our training procedure involves the early stopping callback. This technique involves training as long the loss is decreasing. If the loss is not decreasing for $m$ epochs, training is stopped. The value of $m$ is known as the patience. We set the maximum number of epochs to 150, which is enough to allow the callback to decide when to terminate the training process. We observe that small values of patience, around $m=1..3$ and $m=6$ are sufficient for training on the Ising model and the Rydberg system, respectively. Additionally, we use a learning rate of $\alpha_{lr}=0.001$ for the Ising model, and $\alpha_{lr}=0.0005$ for the Rydberg system. Another hyperparameter is the margin for the contrastive loss, where we use $\alpha=0.4$. We employ the TensorFlow and Keras libraries to implement the network, training, and callbacks.

\subsection{Phase Boundaries from Siamese Networks}
\label{sec:SNNphases}
\subsubsection{Supervised Learning of Phase Transitions:}

Since the proposal to use supervised machine learning algorithms for calculating phase diagrams, the most powerful method still remains using feed-forward neural networks for the binary classification of Monte-Carlo samples \cite{Carrasquilla2017}. In this case a neural network is trained on configurations from known parts of the phase diagram labelled by their phase. By denoting the phases with binary labels $y\in[0,1]$, a neural network $f$ is trained to predict the phase of a configuration $S$

\begin{align}
    f(S) = \left\{
    \begin{array}{l}
       \text{ Phase A} \\
       \text{ Phase B}\\
    \end{array} \right.
    . 
\end{align}

After training, this neural network is then applied to samples from unknown parts of the phase diagram. Since these networks intrinsically learn the underlying physical features characterizing the phases like order parameters and other thermodynamic quantities \cite{Wetzel2017a}, the predictions of the neural networks flip from one label to the other at the position of the phase transition. 

The principle that guides us through the development of an unsupervised Siamese network-based scheme for phase recognition is to leverage the power of neural networks for phase classification in the supervised setting. This is done by reformulating the task of predicting phases by a neural network. A Siamese neural network takes a pair of input configurations and predicts if they are similar or different with respect to a metric imposed by the objective function during training. 

\begin{align}
    F(S_1,S_2) = \left\{
    \begin{array}{l }
       \text{ same Phase } \\
      \text{ different Phase }\\
    \end{array} \right.
\end{align}

In this formulation a SNN can be used as a supervised algorithm for multi-phase recognition. In order to do unsupervised learning of phases with SNNs the training data cannot be supplied with phase labels. In order to enable a SNN to learn from unlabelled data we need to understand how data affects the gradients while training neural networks.

\subsubsection{Gradient Manipulation:}

To develop an unsupervised framework, it is important to understand how gradients update the neural network. Let us discuss this at the example of a general neural network $h$ trained in a supervised setting to minimize the mean square error loss function on a labelled data set $(X,Y)$. The discussion can be extended to any network in this manuscript. The effect of a single training example $(x,y)\in(X,Y)$ on the loss function is

\begin{align}
L(h(x),y)=(h(x)-y)^2
\end{align}
The neural network switches its prediction at the decision boundary 
\begin{align}
    h(x)  \left\{
    \begin{array}{l l}
      <0.5 & \text{ class A} \\
      >0.5 & \text{ class B}\\
    \end{array} \right.
\end{align}

Neural networks are trained using backpropagation of gradients. Let us focus on the gradient signal on an example weight $w$ out of the millions of parameters characterizing a neural network. Let us further assume we have two identical training configurations $x$ with opposite labels labels $y\in[0,1]$. Each update step invloves the product of the learning rate $\eta$ and the inverse of the derivative of the loss function with respect to $w$:
\begin{align}
w_{new}= - \eta \ \del_w L(h(x),y) 
\end{align}
The update depends on the label $y$:
\begin{align}
\del_w L=(h(x)-y)^2\lvert_{f(x)=0.5} = \left\{
    \begin{array}{l l}
      \del_w h(x) &y=0 \\
     -\del_w h(x) &y=1 \\
    \end{array} \right.
\end{align}

Thus, by supplying the neural network with two similar training samples, but opposite labels in the same training step, their effect on the weights of the neural networks would approximately cancel each other out. While this combined training signal forces the neural network to be more uncertain $h(x)\rightarrow 0.5$, it will never change the prediction itself.

\subsubsection{Unsupervised Learning of Phase Transitions:}

In each update step the Siamese network is trained on triplets $(S_a,S_+,S_-)$, where $S_a$ is called anchor, $S_+$ the positive comparison, and $S_-$ the negative comparison. In an unsupervised setting we do not have the true phase labels at hand. However, we know that two samples from the exact same point on the phase diagram must have the same phase. Thus, we create training batches where $S_a$ and $S_+$ are from the same coordinates in the phase diagram, while $S_-$ is sampled randomly from anywhere in the phase diagram.

Building on the discussion of gradient signals: If $S_+$ and $S_-$ stem from the same phase the training signals should approximately cancel each other out, such that the remaining noise is subleading compared to the signal that is obtained when $S_+$ and $S_-$ are from different phases. In a physical context, the noise might stem from thermal fluctuations, and the leading signal from relevant thermodynamical quantities. 

Since there is still no comprehensive theory on neural network training dynamics, the above discussion lacks in mathematical rigor. Hence, in line with all other machine learning based phase recognition methods, the only way to convince ourselves of the capabilities of the proposed method is an empirical study by applying the method to physical systems.

\section{Results}
For the purpose of calculating phase diagrams, we train SNNs as outlined in the previous paragraphs for both the stacked Ising model and the Rydberg atom array. We create training triplets $(S_a,S_+,S_-)$ containing a randomly sampled anchor configuration $S_a$, a positive configuration $S_+$ sampled from the same point in the phase diagram and a negative configuration $S_-$ sampled from any other point in the phase diagram. After having successfully trained a SNN it is employed to perform pairwise comparisons on configurations along a specified one-dimensional slice through the phase diagram.

\subsection{Adjacency Comparisons}
\begin{figure}[htb!]
\begin{center}
\includegraphics[width=1\textwidth]{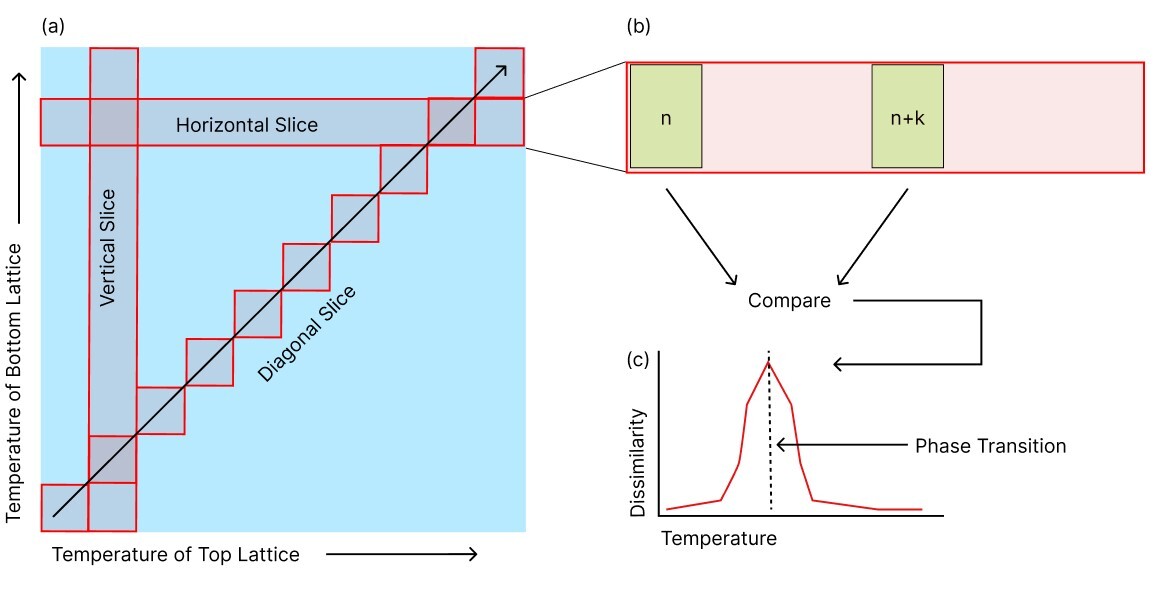}\\
\caption{(a) The horizontal and vertical slices extend through the lattice, while the diagonal slice pierces through the center of the lattice. (b) Within each slice, we compare configurations at indices $n$ and $n+k$ in order to (c) evaluate their dissimilarity. The temperature value at which the peak is situated in (c) corresponds to the predicted phase transition temperature.}
\label{fig:comparisonScheme}
\end{center}
\end{figure}
In this scheme, a configuration corresponding to a point in the phase diagram is compared to its neighbors within a certain distance. At the example of a single Ising model, this means a configuration from temperature $T_n$ is compared to a configuration at $T_{n+k}$, where $n$ is the temperature list index and $k$ is a constant shift. The value of $k$ can be treated as a hyperparameter. A comparison between configurations at $T_n$ and $T_{n+k}$ is assigned a temperature of $\frac{1}{2}(T_{n}+T_{n+k})$.

Our similarity measure of choice is the \emph{normalized cosine dissimilarity} scaled to a range $[0,1]$, where 1 corresponds to the empirically found maximal dissimilarity and 0 to the empirically determined minimal similarity. The cosine dissimilarity is related to the cosine distance via $1-cos(\theta)$, where
\begin{equation}
    cos(\theta)=\frac{f(S_n)\cdot f(S_{n+k})}{\lVert f(S_n)\rVert \lVert f(S_{n+k})\rVert}
\end{equation}
The cosine distance is equivalent to the euclidean distance when acting on the SNN embedding space normalized to the unit sphere ($\lVert f(S_i)\rVert^2=1$):
\begin{align}
    \lVert f(S_n)- f(S_{n+k})\rVert^2 &=2-2 \ cos(\theta)
\end{align}
Since the neural network $f$ is designed to output positive values $f(S_n)_i\in[0,1]$, the cosine dissimilarity takes on its minimum $1-cos(\theta)=0$ for similar embeddings and its maximum $1-cos(\theta)=1$ if the embeddings are dissimilar.

The result of scanning across a phase transition is depicted schematically in \fig{fig:comparisonScheme}(c). The highest peak will indicate the SNN prediction of the phase transition.

\subsection{Ising Model}
\begin{figure}[htb!]
\begin{center}
\includegraphics[width=1\textwidth]{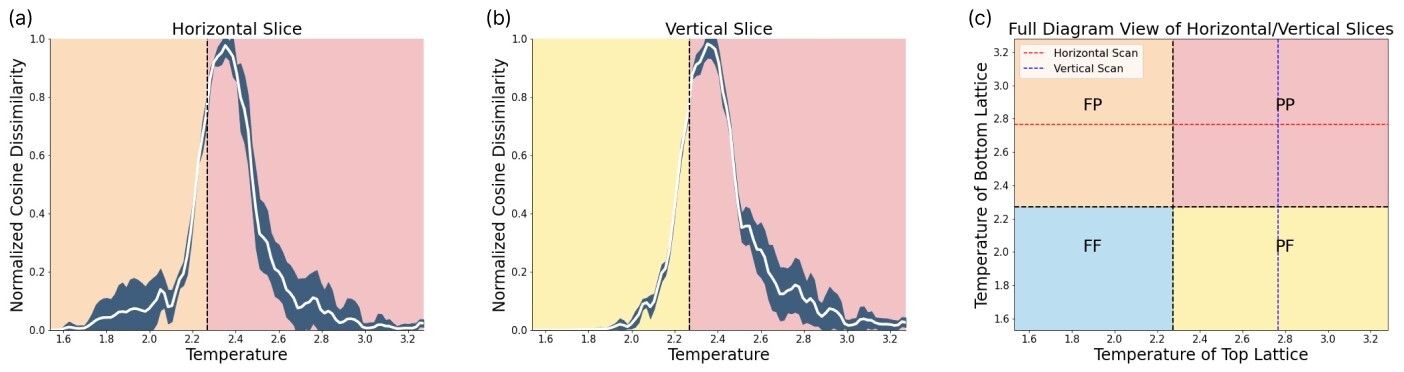}\\
\caption{We use $k=1$ to perform adjacency comparisons across the phase diagram. The blue shading represents the standard deviation over an ensemble of 5 runs. We do not see significant variability in the prediction of the phase transition. These particular slices are taken at (a) $(\cdot,\tilde{T}=2.767)$ and (b) $(T=2.767,\cdot)$. The peaks of both slices predict $T_{c}\approx 2.352$.}
\label{fig:singleHVslice}
\end{center}
\end{figure}
We examine the results of applying SNN unsupervised phase recognition to the $20\times20$ stacked Ising model. Analytically, the phase boundaries of the stacked Ising model in the thermodynamic limit are $(T,\tilde T)=(2.269,\cdot)$ and $(T,\tilde T)=(\cdot,2.269)$. The phase transition temperature is prone to finite size effects as the lattice becomes smaller. If the phase transition would be calculated using the magnetization, finite size effects would distort the phase boundaries to $(T,\tilde T)\approx(2.36,\cdot)$ and $(T,\tilde T)\approx(\cdot,2.36)$; however, it is important to note that the quantities a neural network learns might differ from the magnetization and thus experience different finite size scalings \cite{alexandrou2020critical}. 

\begin{figure}[htb!]
\begin{center}
\includegraphics[width=0.8\textwidth]{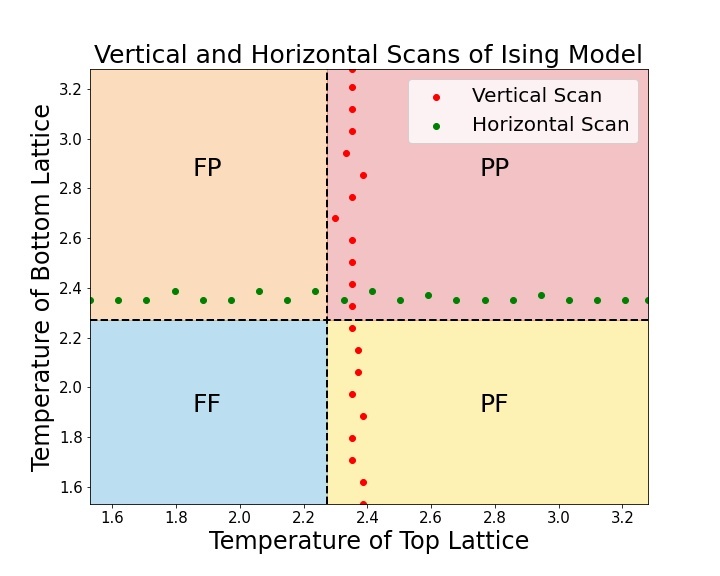}\\
\caption{The phase diagram of the stacked Ising Model contains four regions. Both of the sublattices of the stacked model can be in either the paramagnetic (P) or ferromagnetic (F) phase. We perform adjacency comparisons as in \fig{fig:singleHVslice}(a), (b) for each point in the dotted lines indicating the SNN estimates for the phase transitions. The deviations from the true phase transitions are quantitatively consistent with finite size effects on the magnetization. }
\label{fig:fullIsingSlices}
\end{center}
\end{figure}

In order to calculate the stacked Ising model phase boundaries, we choose to perform vertical and horizontal scans across the phase diagram, where each scan is repeated five times and the standard deviation of this ensemble is displayed as uncertainty. Exemplarily, the results of two of these scans can be found in \fig{fig:singleHVslice}. The location of these scans is depicted in (c). (a) reveals the phase transition from (ferromagnetic, paramagnetic) to (paramagnetic, paramagnetic)  (FP to PP), while (b) displays the phase transition from PF to PP. Collecting 21 vertical and 21 horizontal scans reveals the phase boundaries of the stacked Ising Model, as seen in \fig{fig:fullIsingSlices}. By comparing the horizontal (green dotted line) and vertical scans (red dotted line) with the Ising model phase boundaries in the thermodynamic limit (black dashed lines) one can observe a clear difference. The SNN predicts critical temperatures of $T_{c}\approx 2.352$, consistent with the finite size correction of the magnetization which indicates a phase transition at $T_{c}\approx 2.36$.

\begin{figure}[htb!]
\begin{center}
\includegraphics[width=1\textwidth]{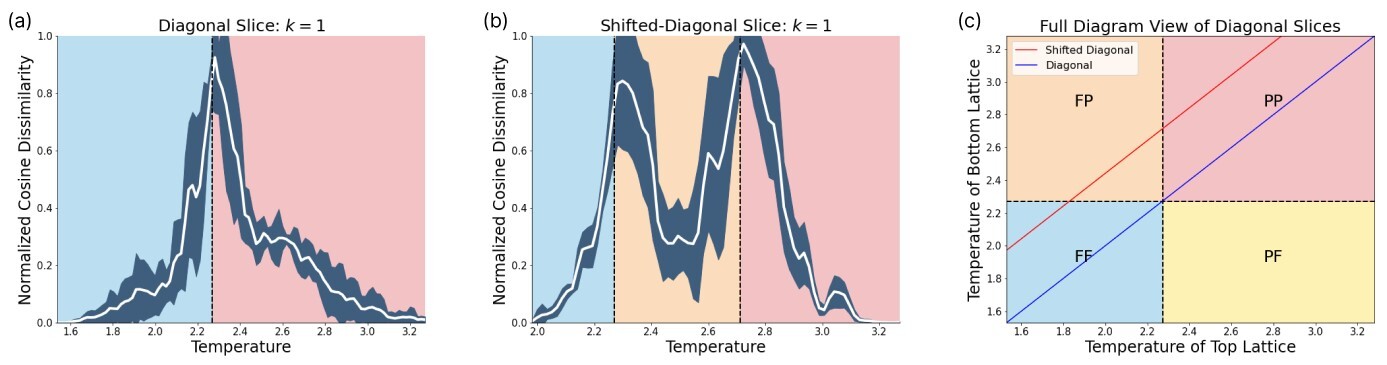}\\
\caption{We perform a two different diagonal scans across phase diagram of the stacked Ising Model. The adjacency comparison plot (a) depicts the result from a scan across the diagonal (blue line in (c)), while (b) depicts the same for the shifted diagonal (red line in (c)). In (c), the diagonal (blue) line only crosses the phase boundaries once in the center (FF to PP), and we correctly observe a single corresponding peak in (a). Likewise, the shifted diagonal line (red) crosses two phase boundaries (from FF to FP, then FP to PP), resulting in two corresponding peaks.}
\label{fig:k1plots}
\end{center}
\end{figure}

In order to reveal the power of our SNN based phase recognition scheme, we have to scan across diagonal slices within the phase diagram that contain more than one phase transition. For this purpose we scan diagonally across the phase diagram as depicted in \fig{fig:k1plots}. The two diagonal scans across the lattice reveal that this technique can identify more than one phase transition. The first diagonal scan is performed along a line through the center of the lattice. In this case, we see a single phase transition, as the network scans directly through the intersection of the vertical and horizontal phase lines. A closer examination of the effect of changing $k$ can be found in \fig{fig:diagonalDirect}. The prediction yields a phase boundary at $(T,\tilde T)_c=(2.28, 2.28)$.

The second diagonal scan we perform is shifted, such that it crosses both the vertical and horizontal phase transition. In this case, the true phase boundaries are cut by the shifted diagonal slice at $(T,\tilde T)_c=(1.827, 2.269)$ and $(T,\tilde T)_c=(2.269, 2.711)$. The network is able to capture both transitions at $(T,\tilde T)_c\approx(1.857, 2.299)$ and $(T,\tilde T)_c\approx(2.281, 2.723)$. A closer examination is found in \fig{fig:shifted}.

% \fig{fig:diagonalDirect}\fig{fig:shifted} \zak{I think we can del these now, right?}

% \fig{fig:IsingEmbedsNonShift}

% \fig{fig:IsingEmbedsShift}

\begin{figure}[htb!]
\begin{center}
\includegraphics[width=1\textwidth]{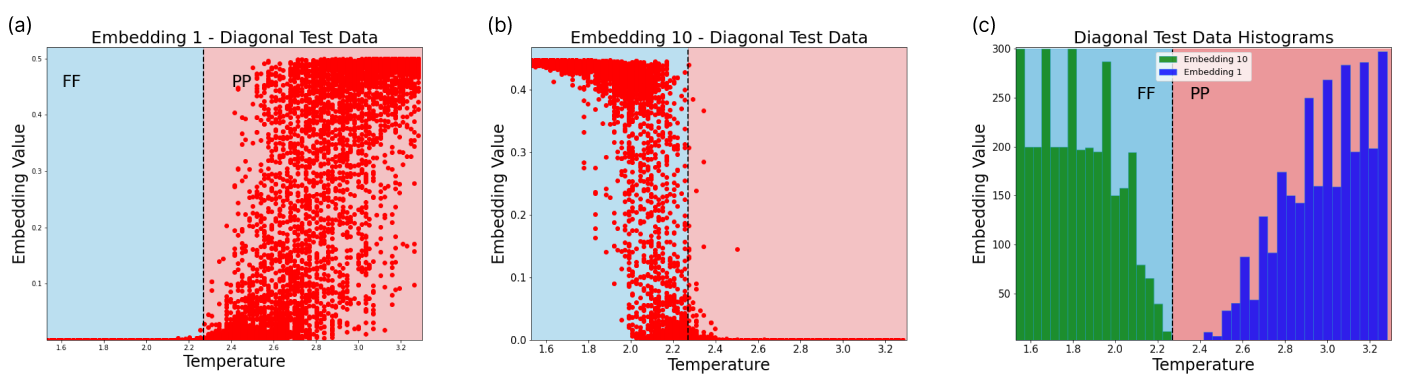}\\
\caption{
Latent space embeddings of the SNN are applied to configurations of the stacked Ising model. In particular, we choose our configurations from the diagonal slice to reveal how the SNN encodes phase information. Of the 10 embedding neurons, some encode the phase information of the diagonal slice. Embedding neurons 1 and 10 are such neurons, distinguishing between the FF and PP phases in the diagonal slice. (a) and (b) depict a separation of activity between the aforesaid regions. (c) counts the number of instances in which a point is above the halfway line in the y-axis of (a) and (b). There is a separation in these histograms between the two phase regions, FF and PP. According to (c), embedding 1 (blue) has a preference for encoding information in the PP region, while embedding 10 (green) encodes the FF region.}
\label{fig:nonshiftEmb}
\end{center}
\end{figure}

Furthermore, the embeddings shed some light on how the neural network encodes the aforesaid phase transitions. In \fig{fig:nonshiftEmb} we see embeddings which encode phase information for the diagonal slice. These embeddings clearly separate between the two underlying phases. Embedding 1 only spikes in the PP phase, while embedding 10 activates at FF regimes. A switch between both embeddings occurs at the phase transition. Other embeddings can be found in \fig{fig:IsingEmbedsNonShift}.
In the case of a single embedding, we may have noise contained in the same embedding neuron where the phase behaviour is captured. With additional embeddings, the noise may be relegated to other embedding neurons, isolating the phase transition information. 

\begin{figure}[htb!]
\begin{center}
\includegraphics[width=1\textwidth]{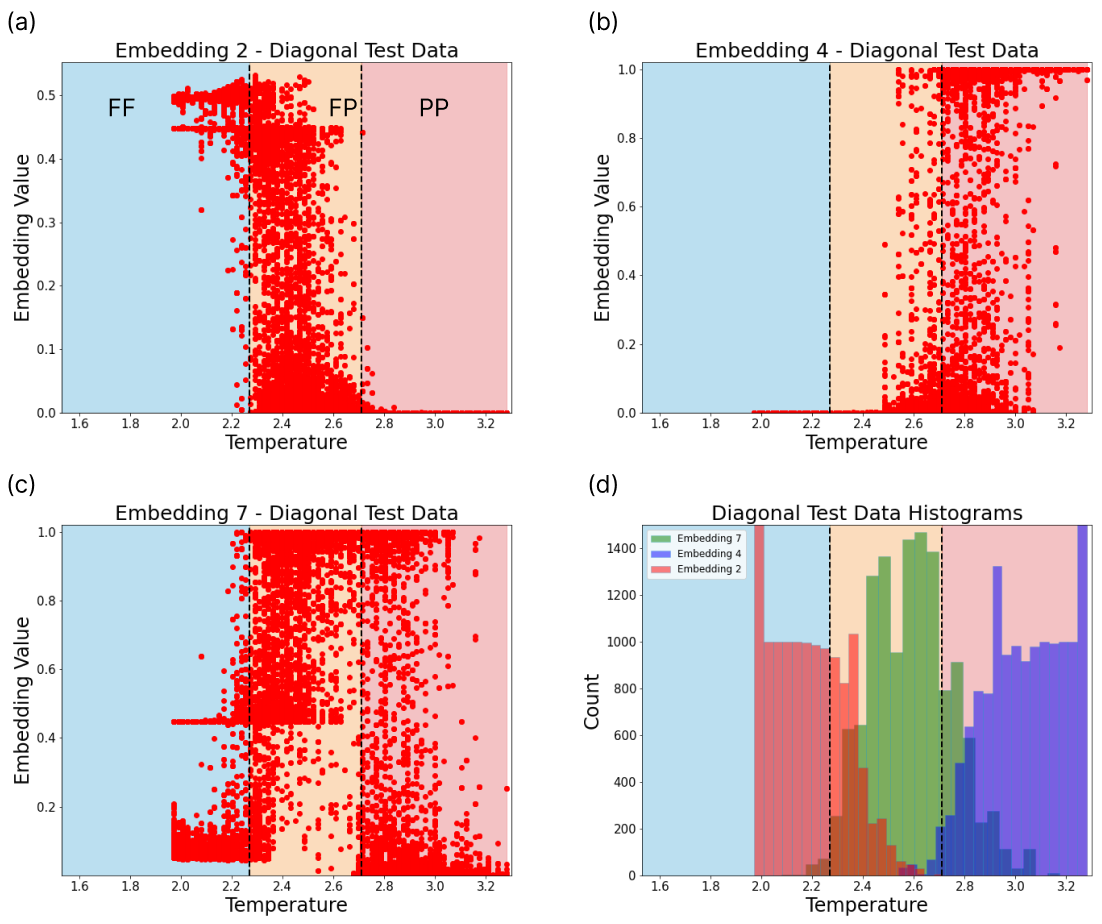}\\
\caption{Latent space embeddings of the SNN are applied to configurations of the stacked Ising model belonging to the shifted diagonal slice, which contains two phase transitions. The embeddings display structures associated with each phase. While (a) appears to encode the FP phase, its corresponding histogram reveals that it encodes the FF phase. The histogram accounts for the density of points beyond the halfway line in (a), indicating that many points are clustered together in the FF phase. (c) similarly shows the encoding of the FP phase, while (b) shows the encoding of the PP phase. The corresponding histograms are colour-coded in (d).}
\label{fig:shiftEmb}
\end{center}
\end{figure}

The embedding space of the shifted diagonal scan is depicted in \fig{fig:shiftEmb}. We observe three crucial behaviours in the embeddings. Each of the three histograms show maximum activity in three different regions. Embedding 2 encodes the FF phase, embedding 4 the PP phase, and embedding 7 encodes configurations from the FP phase. 

\subsection{Rydberg Array}
The Rydberg atom array phase diagram is not as well studied as the Ising model, so we need to compare our SNN phase boundaries to recent papers \cite{ebadi2021,samajdar2020} and evaluate the order parameters on the QMC data ourselves. 

In order to identify the approximate phase boundaries we compute predictors for the phases of interest. 
Each phase corresponds to various peaks in the absolute value of the Fourier transform (FT) of the one-point function:
\begin{equation}\label{eq:oneptfn}
    n(\mathbf{k})=n(k_x, k_y) = \abs{\frac{1}{\sqrt{N}}\sum_j n_j \exp(i \mathbf{k}\cdot\mathbf{r}_j)}
\end{equation}
where $n_j$ is the Rydberg occupation of the $j$th site.
Furthermore, we symmetrize the FT by averaging over permutations of the momentum axes:
\begin{equation}\label{eq:symft}
    \mathcal{F}(k_x, k_y) = \frac{1}{2}\left[n(k_x, k_y) + n(k_y, k_x)\right]
\end{equation}
The peaks occur for the checkerboard phase at $(\pi, \pi)$, for the striated phase at $(\pi, 0)$ and $(\pi, \pi)$, and for the star phase at $(\pi, 0)$, and $(\pi/2, \pi)$\cite{samajdar2020}. We compute, for each point in parameter space $(R_b, \delta)$, the symmetrized FT at these specific momenta averaged over the full 1200 sample data set given to the SNNs.

\subsubsection{Testing}
% \fig{fig:rydbergPlots} slices

% \fig{fig:rydbergPDs} phase diags
% \fig{fig:rydberg1}
% \fig{fig:rydberg2}

\begin{center}
\begin{figure}[htb!]
\includegraphics[width=1\textwidth]{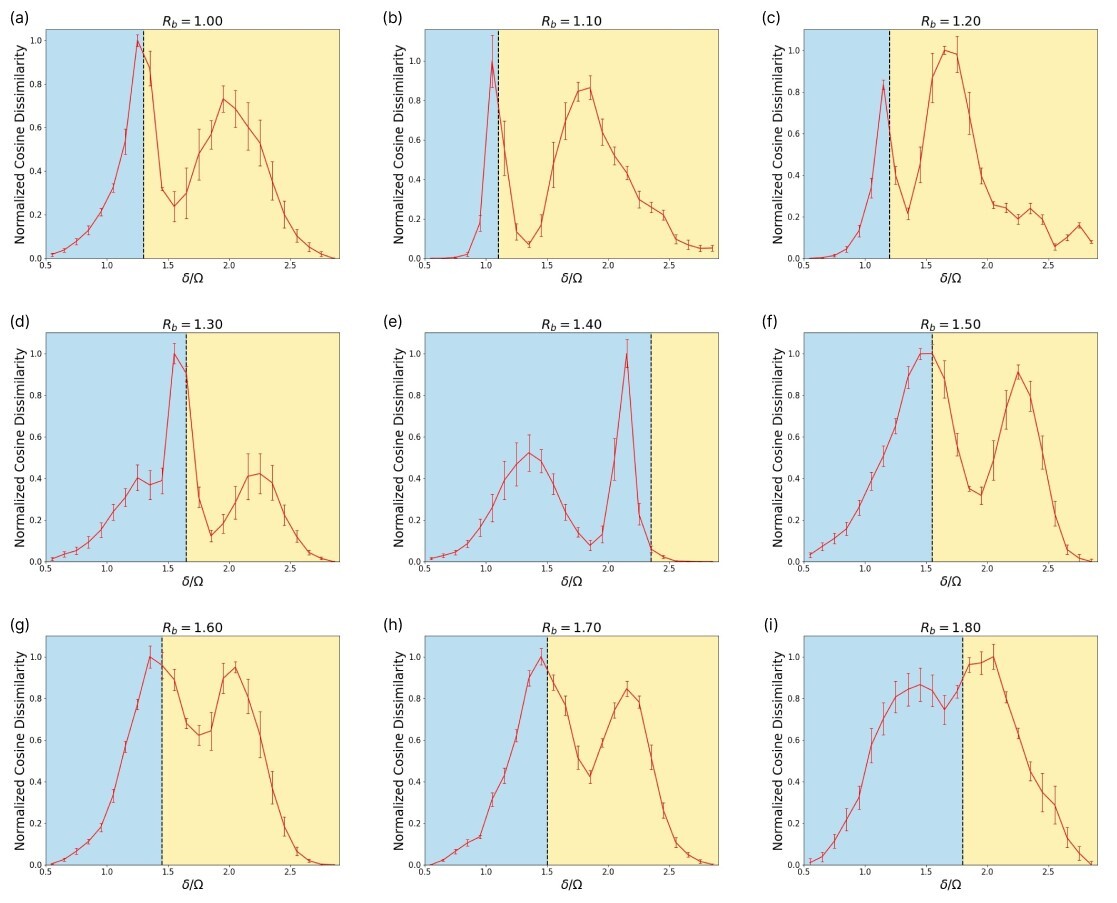}\\
\caption{We apply the SNN phase recognition method to the Rydberg array in the region $R_b \in \{1.0, \ldots, 1.8\}$ and $\delta \in \{0.5, \ldots, 2.9\}$. We scan along each horizontal slice by choosing a fixed $R_b$, and traversing along all $\delta \in \{0.5, \ldots, 2.9\}$. We choose $k=1$ to perform adjacency comparisons within these scans. The dissimilarity peaks when configurations at $\delta_i$ and $\delta_{i+1}$ are of different phases.}

% produce pairs by choosing configurations $S_i, S_{i+1}$ at $(\delta_i, \delta_{i+1})$ (i.e., $k=1$). Each plot (a)-(i) reveals a peak where $S_i$ is most dissimilar to $S_{i+1}$. }

\label{fig:rydbergPlots}
\end{figure}
\end{center}

The Rydberg atom array phase diagram is examined in regimes where the checkerboard, striated and star phases are present. Thus, guided  by \cite{ebadi2021} we focus the region $R_b \in \{1.0, \ldots, 1.8\}$ and $\delta \in \{0.5, \ldots, 2.9\}$. The training of the SNN on Rydberg QMC data is performed similar to the Ising model on each horizontal and vertical slice. Inference is performed by adjacency comparisons with step-size $k = 1$. Because the Rydberg phase diagram is coarser than the Ising model phase diagram (25 points between $\delta = 0.5$ to $\delta = 2.90$ for the Rydberg array, compared to 100 points between $T = 1.53$ and $T = 3.28$ for the Ising model), there is no reason to use $k > 1$. 

\begin{figure}[htb!]
\begin{center}
\includegraphics[width=1\textwidth]{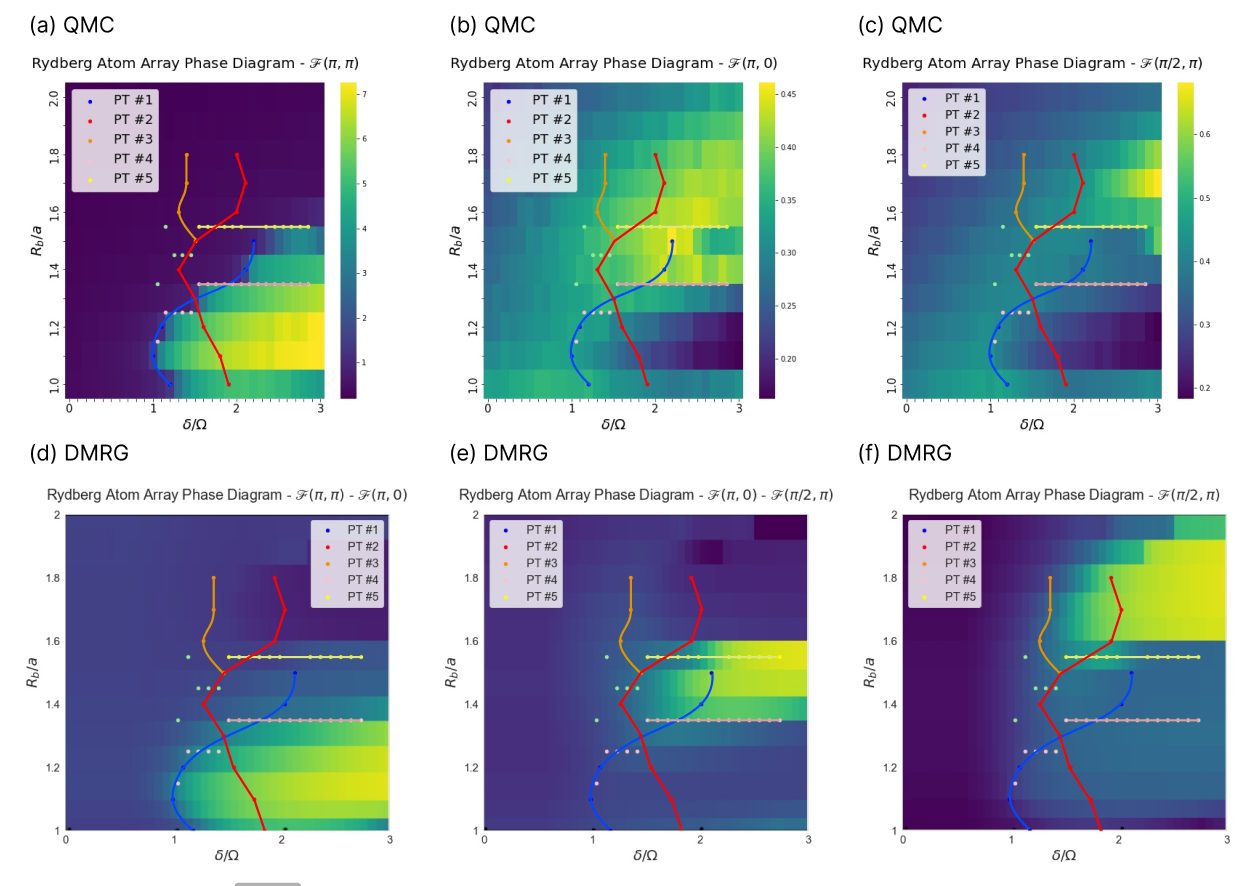}\\
\caption{Phase transition lines are superposed on the Rydberg phase diagram. These lines result from connecting the peaks from \fig{fig:rydbergPlots} and \fig{fig:delta} in the smoothest possible way. Doing so results in three distinct phase boundaries. In (a)-(c) we overlay these phase lines on the diagram generated via our QMC simulation, whereas (d)-(f) use the phase diagram produced by \cite{ebadi2021}.}
\label{fig:rydbergPDs}
\end{center}
\end{figure}

All horizontal and vertical scans depicted in \fig{fig:rydbergPlots} and \fig{fig:delta}, respectively, are the result of an ensemble average over 5 different runs per slice. The error bars represent the standard deviation between runs. The vertical lines in each plot in \fig{fig:rydbergPlots} represent the approximate phase transition at each $R_b$ based on \cite{samajdar2020}. Our observations differ from their result in the sense that we consistently observe two phase transitions in all slices (although some of these phase transition peaks are weak, as shown in \fig{fig:rydbergPlots}). However, we also confirm that in all cases the SNN predicts a phase transition that coincides with the results from \cite{samajdar2020}. There is only a slight deviation at $R_b=1.4$ in the horizontal scanning direction. Furthermore, the results in \fig{fig:delta} reveal the striated phase in between the star and checkerboard phases, which is also consistent with \cite{samajdar2020}.

\begin{center}
\begin{figure}[htb!]
\includegraphics[width=1\textwidth]{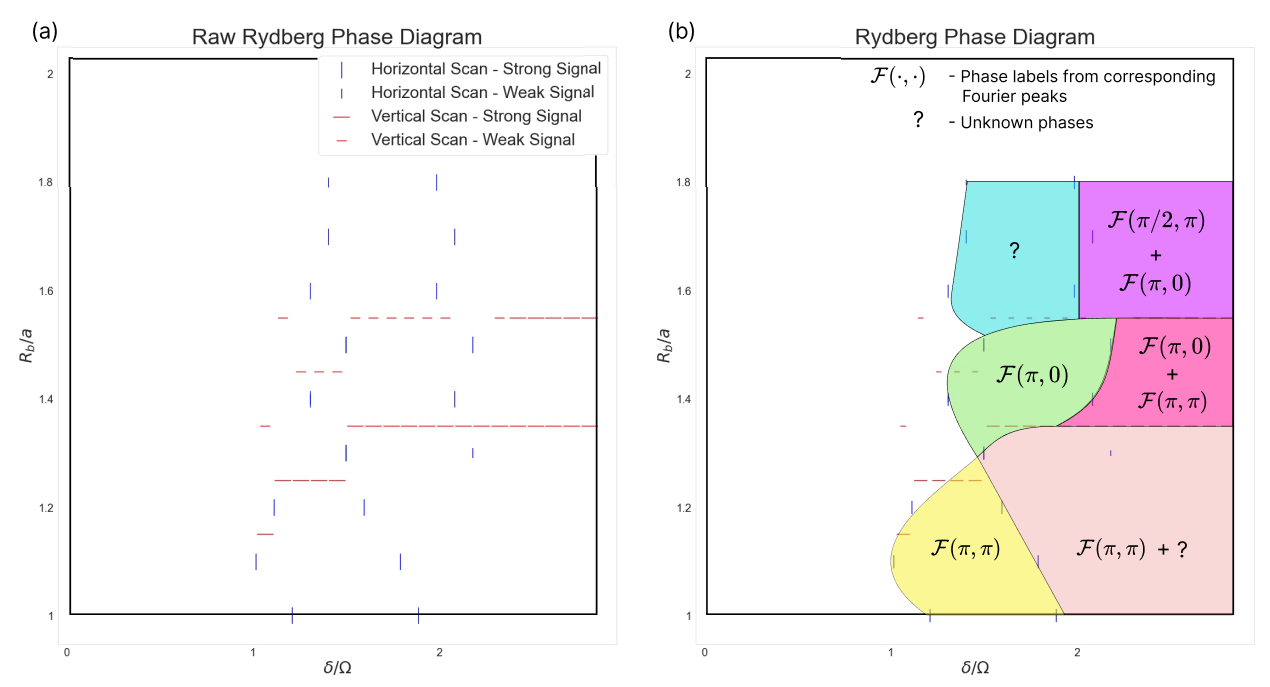}\\
\caption{SNN phase recognition method applied to the Rydberg array in the region $R_b \in \{1.0, \ldots, 1.8\}$ and $\delta \in \{0.5, \ldots, 2.9\}$. (a) Raw results of SNN scans along horizontal and vertical slices collected from \fig{fig:rydbergPlots} and \fig{fig:delta}. Strong signals correspond to clearly visible phase transition signals, while weak signals are only considered phase transition indicators if they can be consistently identified in neighboring slices. (b) Regions in the phase diagram are separated by connecting markers in (a) as smoothly as possible, they are labelled according to the Fourier Modes characterizing the phases in \fig{fig:rydbergPDs}(a)-(c).}

\label{fig:completePD}
\end{figure}
\end{center}
In figures \fig{fig:rydbergPDs} (a)-(c) we compare the SNN predictions with the results from evaluating order parameters on our QMC data. For each the checkerboard phase, the star phase, and the striated phase we create separate plots. The QMC data is chosen as the background, while in the foreground we connect phase transition signals from the SNN in the smoothest way possible in the form of blue, red, orange, pink and yellow lines. We first observe a perfect agreement of the checkerboard phase transitions seen in (a). The striated and star phase in figures (b) and (c) are very elusive; however, the SNN tends to capture clearer phase information than what is suggested by the evaluation of the order parameters. The red line captures the striated phase for $R_b>1.3$. There are two distinct differences between SNN and QMC order parameter results:  There is no QMC order parameter signal to explain the orange line. Further, the red line continues well within the checkerboard phase, where none of the three order parameters signal a phase transition.

We also compare our results to those obtained by \cite{ebadi2021} in \fig{fig:rydbergPDs} (d)-(f). Again the SNN prediction of the checkerboard phase transition is in good agreement with the DMRG results. Further, the red line for $1.3<R_b<1.6$ is similar to their striated phase boundary. The pink and yellow lines also bound the striated phase. In addition, our orange line corresponds very well to the DMRG star phase. The DMRG star phase is much larger than the QMC star phase. However, the SNN results should mimic what is present in the QMC data. This contradiction might be resolved by two different explanations: 1) the neural network is able to extract features which are a stronger indicator for the star phase than Fourier modes. 2) there is another phase present in the QMC data that is responsible for the orange phase boundary, \cite{samajdar2020} suggests possible candidates in form of rhombic, banded or staggered order.

\begin{center}
\begin{figure}[htb!]
\includegraphics[width=\textwidth]{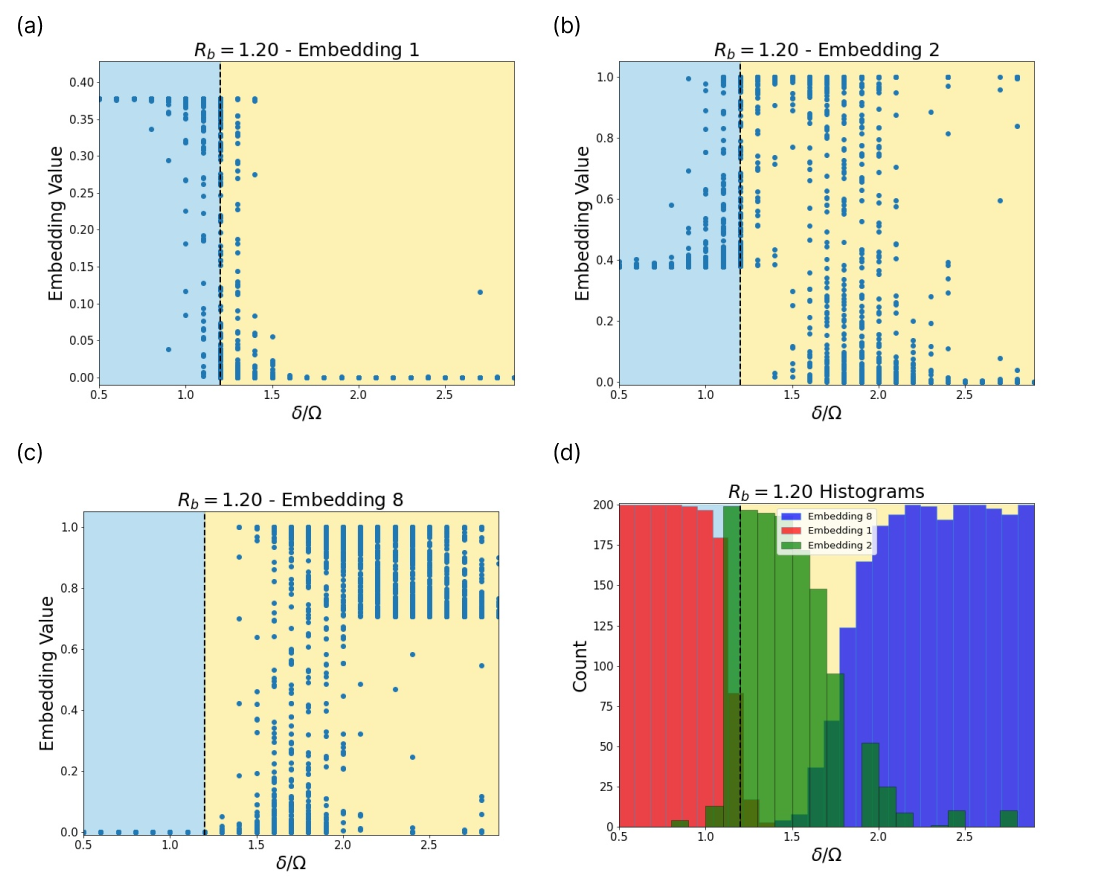}\\
\caption{Latent space embeddings of the SNN are applied to configurations of the Rydberg system belonging to the horizontal slice along $R_b=1.20$ in the checkerboard region. Of the 10 embedding neurons, some encode phase information along $R_b=1.20$. Embeddings 1, 2 and 8 are such neurons, revealing a phase transition near $\delta=1.20$, which is marked by the black dashed line. The phase information is markedly partitioned by this line, as can be seen in (a) and the corresponding red embedding in (d). This phase transition corresponds to the first peak in \fig{fig:rydbergPlots}(c). The second peak in \fig{fig:rydbergPlots}(c) is located near $R_b = 1.70$. Embedding 8 in (c) and the blue histogram in (d) reveal a phase transition near $\delta=1.70$. (b) shows the encoding of an intermediate phase, with the corresponding histogram in (d) coloured green.}

\label{fig:rydberg1}
\end{figure}
\end{center}

The full independent SNN prediction of the Rydberg Array phase diagram is constructed in \fig{fig:completePD}, where (a) depicts the raw peaks of the horizontal and vertical scans in \fig{fig:rydbergPlots} and \fig{fig:delta}, respectively. The vertical scan slices between $\delta \in \{0.50, \ldots, 1.00\}$ do not reveal any explicit phase transition, as expected, and are therefore not included in the phase diagram. \fig{fig:completePD}(b) is constructed by connecting markers as smoothly as possible. This results in 7 phase regions, labelled according to \fig{fig:rydbergPDs}. 

\fig{fig:rydberg1} provides insight into how the SNN encodes the phase information in the Rydberg system. Certain embeddings correspond to very specific regions in the phase diagram. Embeddings 1 and 2 separate in the vicinity of the first phase transition, and where embeddings 2 and 8 separate, the SNN indicates a second phase transition.

It is surprising that the SNN is able to reveal clearer phase information from the QMC data than what a direct evaluation of order parameters might indicate. The reason for this might be that the neural network is able to calculate other thermodynamic quantities which are relevant to describe the underlying physics, such as energies or susceptibilities. Further, the neural network might be able to deal better with domain boundaries. This observation encourages us to predict that neural networks might be able to reveal phases where conventional order parameter evaluations on MC configurations have trouble doing so. In that sense the SNN predicts the occurrence of 2 phases that are not evident in the QMC order parameter analysis, the blue and orange regions in \fig{fig:completePD}. These findings guide us to examine these regions closer with a keen eye on revealing previously unknown physics.

\section{Conclusion}
We have introduced a Siamese neural network (SNN) based method to detect phase boundaries in an unsupervised manner. This method does not require any physical knowledge about the nature or existence of the underlying phases. SNN based phase detection shares the power of a) feed forward neural networks, which have been the most powerful machine learning algorithm to be applied to reveal physical phases, and b) certain unsupervised methods which can learn multiple phases without knowing about their existence. This method is shown to reproduce phase diagrams when trained on Monte-Carlo configurations of the corresponding physical system. In our case we introduced the method at the example of a model consisting of two stacked Ising Models exhibiting a phase diagram of four phases. Further, we used this method to calculate the phase diagram of a Rydberg atom array which is to the most extent consistent with prior results, and shows additional signatures of unknown and coexistence phases. Futher, in some regimes the SNN tends to be better at picking up phase information than order parameters applied to QMC data. As typical for neural network based phase recognition schemes, we do not have insight on what features a neural network is learning in order to calculate the phase boundaries. These quantities have been shown to be related to order parameters and other physically relevant quantities. Explicitly revealing them is difficult, but possible using methods like \cite{Wetzel2017a,miles2021correlator}.

With this work we have contributed to the zoo of machine learning methods for phase diagrams. While it remains to be seen if one of these methods will reveal a completely new unknown phase, we believe SNN based phase detection has the features of a top contender with its ability to detect multiple phases in an unsupervised manner.

\section{Acknowledgements}
We thank Roger Melko for helpful discussions. We thank the National Research Council of Canada for their partnership with Perimeter on the PIQuIL. Research at Perimeter Institute is supported in part by the Government of Canada through the Department of Innovation, Science and Economic Development Canada and by the Province of Ontario through the Ministry of Colleges and Universities.

\appendix

\section{Normalized Cosine Dissimilarity}

\begin{figure}[htb!]
\begin{center}
\includegraphics[width=1\textwidth]{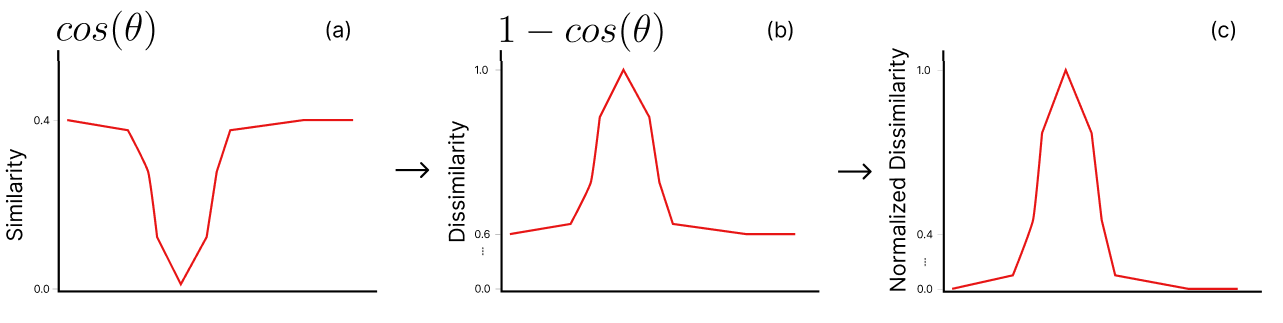}\\
\caption{(a) We begin with the standard cosine similarity plot. (b) We then compute $1-cos(\theta)$ to obtain the cosine dissimilarity. (c) We normalize to guarantee that the plot ranges from 0 to 1. 
% Because $f(S_n)$ is not guaranteed to be orthogonal to $f(S_{n+k})$ for some $n$, we may not obtain $1-cos(\theta_{n,n+k})=1$ for a particular comparison in our plot. Thus, we normalize to achieve a plot containing values in [0, 1].
}
\label{fig:cosine}
\end{center}
\end{figure}

In this section we display the calculation of the cosine dissimilarity in \fig{fig:cosine}. 

\section{Training}

\begin{figure}[htb!]
\begin{center}
\includegraphics[width=1\textwidth]{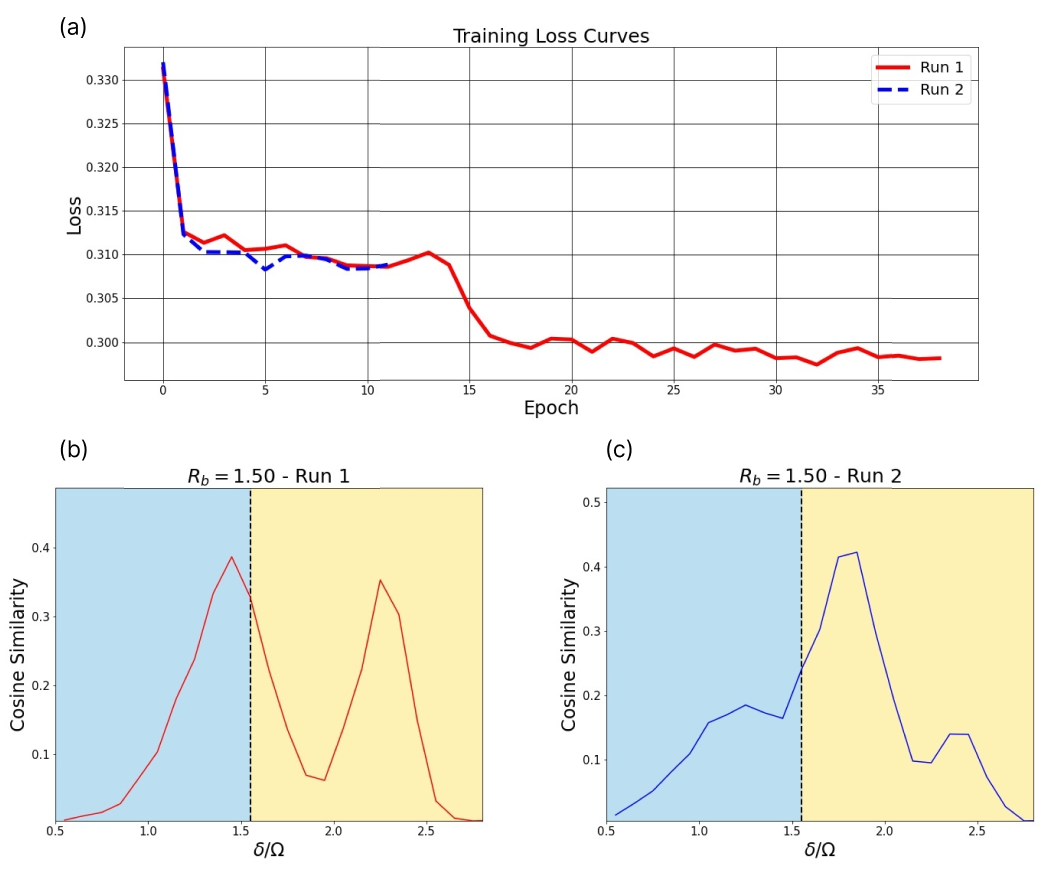}\\
\caption{We train on Rydberg array configurations belonging to the slice $R_b=1.50$, employing early stopping. (a) The blue loss curve reveals that a single peak, shown in (c), occurs when training falls prey to local minima. Training ceases at around 12 epochs in this case due to the early stopping callback. Furthermore, we observe that the red loss curve succeeds in escaping this minimum, and converges to a lower loss. This lower loss corresponds to dual peaks, shown in (b).}
\label{fig:LossCurve}
\end{center}
\end{figure}

\subsection{General Training Dynamics}

We observe that training is prone to getting stuck in local minima. For this reason, we may rerun the training phase several times for certain slices. In particular, the diagonal slices (both the shifted diagonal and the direct diagonal) are rerun if their adjacency comparisons exhibit unexpected/sub-optimal behaviour. One may use the loss curve to identify such instances. For instance, one run may exhibit a single phase transition, while the other may exhibit 2. We opt to keep the result with the lower loss. This scenario is depicted in \fig{fig:LossCurve}. The single peak result in (c) corresponds to a higher loss. We identify this as a local minimum. The red curve in (a) corresponds to the plot (b). While this run also gets stuck in a local minimum, it is able to escape upon further training. As such, we keep (b) as our result. 

\subsection{Ising Model - Training on the Diagonal Slice}

The diagonal slice is shown in \fig{fig:comparisonScheme}(a) and \fig{fig:k1plots}(c). Of the 92 Ising configurations produced at each temperature, 50 are used in the training set, and 42 in the testing set. Then, we use these to produce 100 stacked configurations at random for both the training and testing sets. For example, we choose two random configurations out of the set of 50 for the training set, and overlay them to produce a single stacked configuration. 

% \zak{Loss function we use}

\subsection{Ising Model - Training on the Shifted Diagonal Slice}
 
% \zak{change shift to be in x}

The shifted diagonal slice is shown in \fig{fig:k1plots}(c). Here, we shift the slice by 25 temperature points vertically in order to cross two phase transitions. Of the 92 Ising configurations produced at each temperature, 70 are used in the training set, and 22 in the testing set. Once again, we use these to produce stacked configurations at random for both the training and testing sets. However, this time we produce 3500 stacked configurations for the training set, and 500 for the testing set. Because the diagonal slice does not contain many points from any of the three phases it crosses, we increase the amount of training data to compensate. 

\subsection{Ising Model - Training on the Horizontal/Vertical Slices}
Of the 92 Ising configurations produced at each temperature, 42 are used in the training set, and 50 in the testing set. As with the diagonal slice, we then use these to produce 100 stacked configurations at random for both the training and testing sets.

\subsection{Training the Rydberg System}

Training on the Rydberg system is performed similar to the Ising model, i.e. slice-wise. We train over $R_b \in \{1.00, \ldots, 1.80\}$ and $\delta \in \{0.5, \ldots, 2.9\}$, with 1000 configurations per $\delta$-step. 800 are used for training, and the remaining 200 for testing. 
\newpage

\section{Additional Embeddings}

This section presents the remaining embeddings for the Ising Model's diagonal and shifted-diagonal scans, as well as the $R_b=1.20$ horizontal scan. 

\subsection{Ising Model - Diagonal Embeddings}

\begin{figure}[htb!]
\begin{center}
\includegraphics[width=\textwidth]{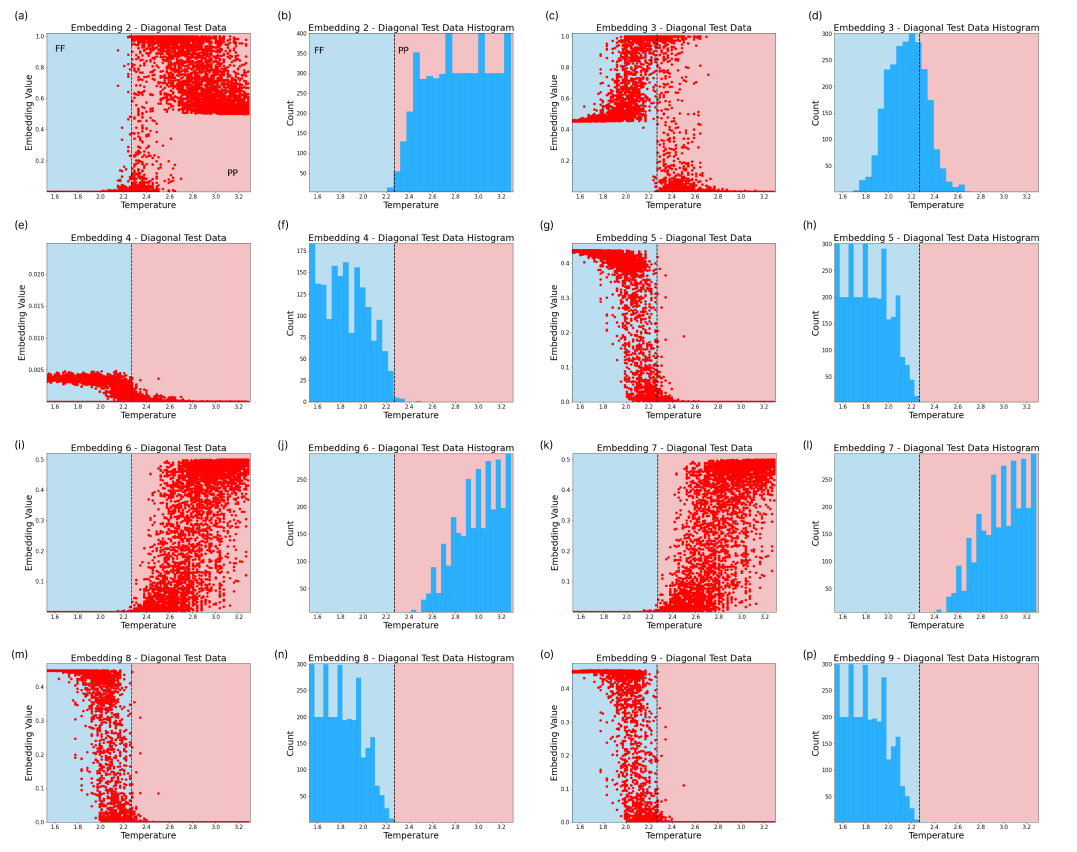}\\
\caption{Latent space embeddings of the SNN are applied to configurations of the stacked Ising Model belonging to the diagonal slice. This slice contains a single phase transition between FF and PP, as depicted in \fig{fig:k1plots}(c). Some neurons belonging to these embeddings encode phase information, albeit redundantly. For instance, (j) and (l) both encode the PP phase, but almost identically. Other embeddings appear unique, but still encode information about the same phase transition. (j) and (m) encode the PP and FF phases, respectively, but the both provide insight into the same phase transition depicted by the black dashed line.}
\label{fig:IsingEmbedsNonShift}
\end{center}
\end{figure}
\newpage

\subsection{Ising Model - Shifted Diagonal Embeddings}

\begin{figure}[htb!]
\begin{center}
\includegraphics[width=\textwidth]{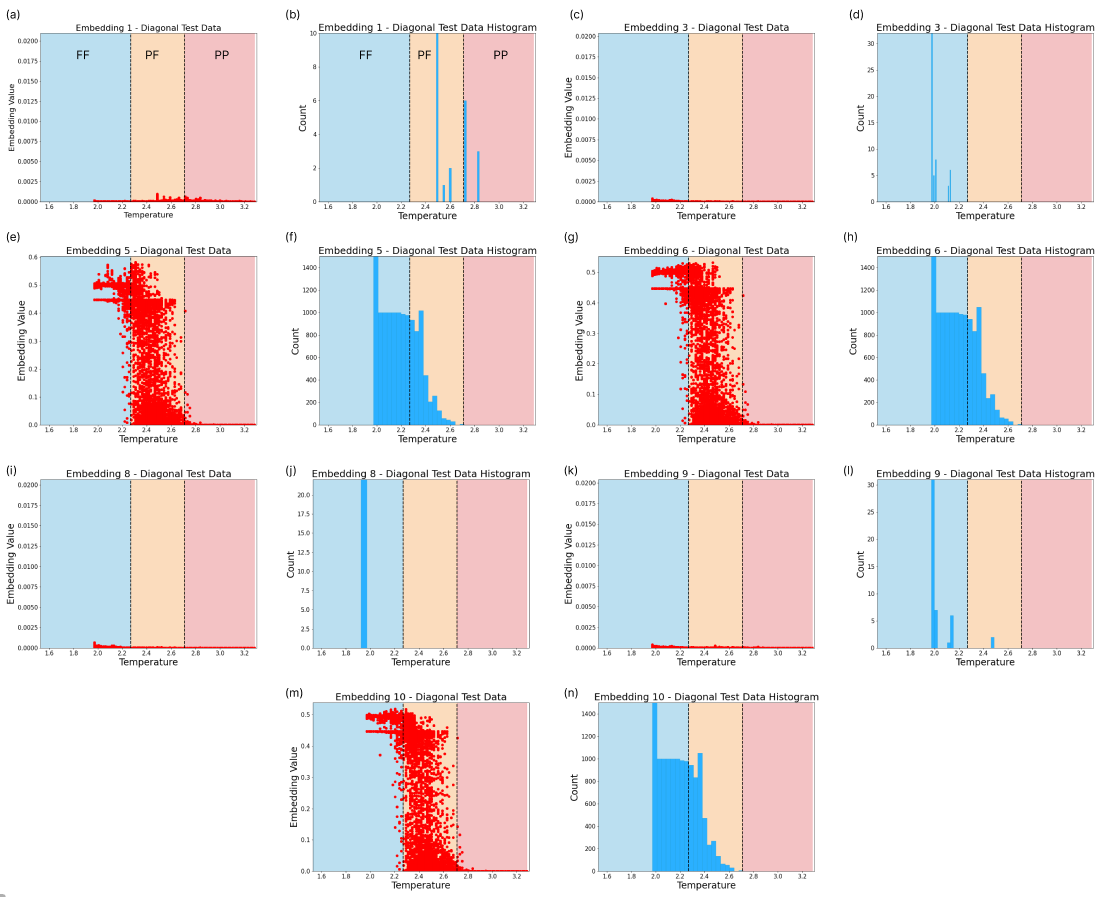}\\
\caption{Latent space embeddings of the SNN are applied to configurations of the stacked Ising Model belonging to the shifted diagonal slice. This slice contains three regions - FF, PF, and PP, as depicted in (a), (b). Some neurons belonging to these embeddings encode phase information, albeit redundantly, while other neurons do not appear to encode anything at all. For instance, (e) and (g) both encode the FF phase, but almost identically. On the other hand, (a), (c), (i), and (k) do not encode any phase information.}
\label{fig:IsingEmbedsShift}
\end{center}
\end{figure}
\newpage

\subsection{Rydberg System}

\begin{figure}[htb!]
\begin{center}
\includegraphics[width=\textwidth]{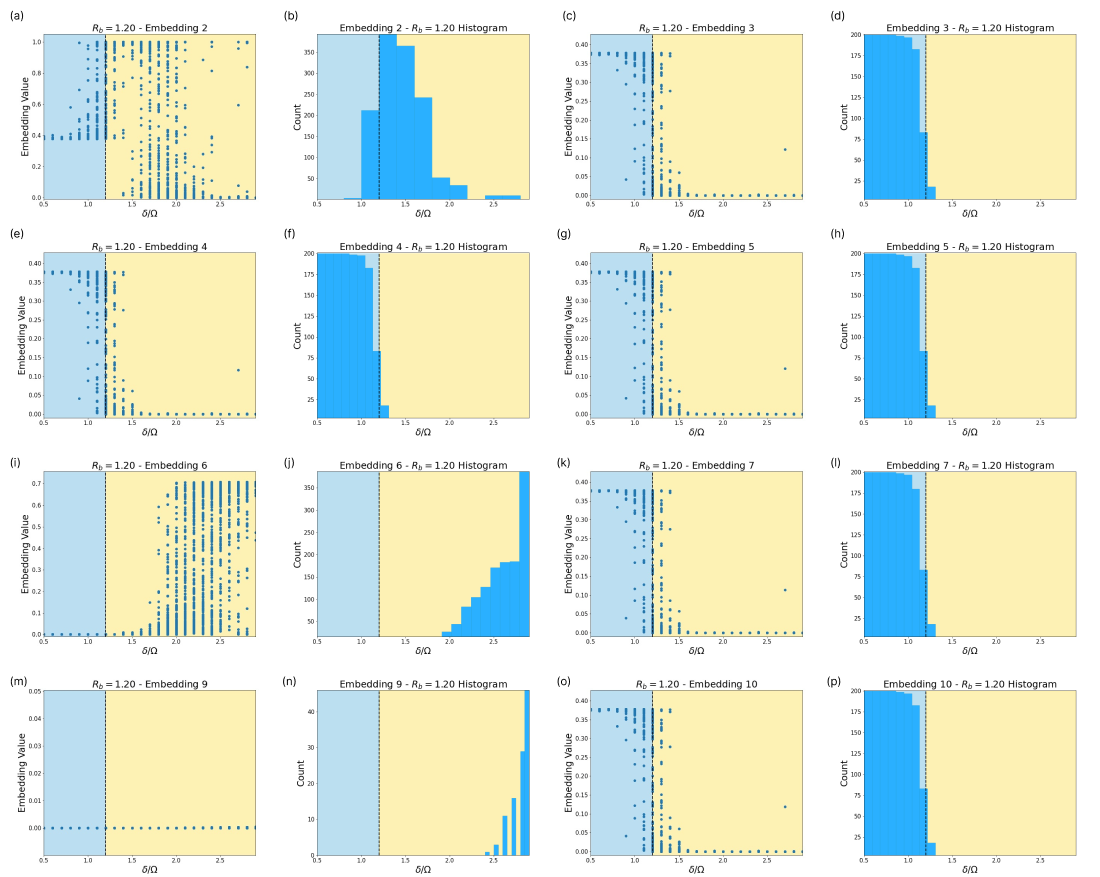}\\
\caption{Latent space embeddings of the SNN are applied to configurations of the Rydberg system belonging to the horizontal slice along $R_b=1.20$ in the checkerboard region. Most neurons encode redundant information. For instance, (a) and (c) encode information related to the same phase, while (c) and (f) exhibit identical phase structures. We also observe that embedding 9 in (m) encodes no phase information.}
\label{fig:RydbergAdd1}
\end{center}
\end{figure}
\newpage

\begin{figure}[htb!]
\begin{center}
\includegraphics[width=\textwidth]{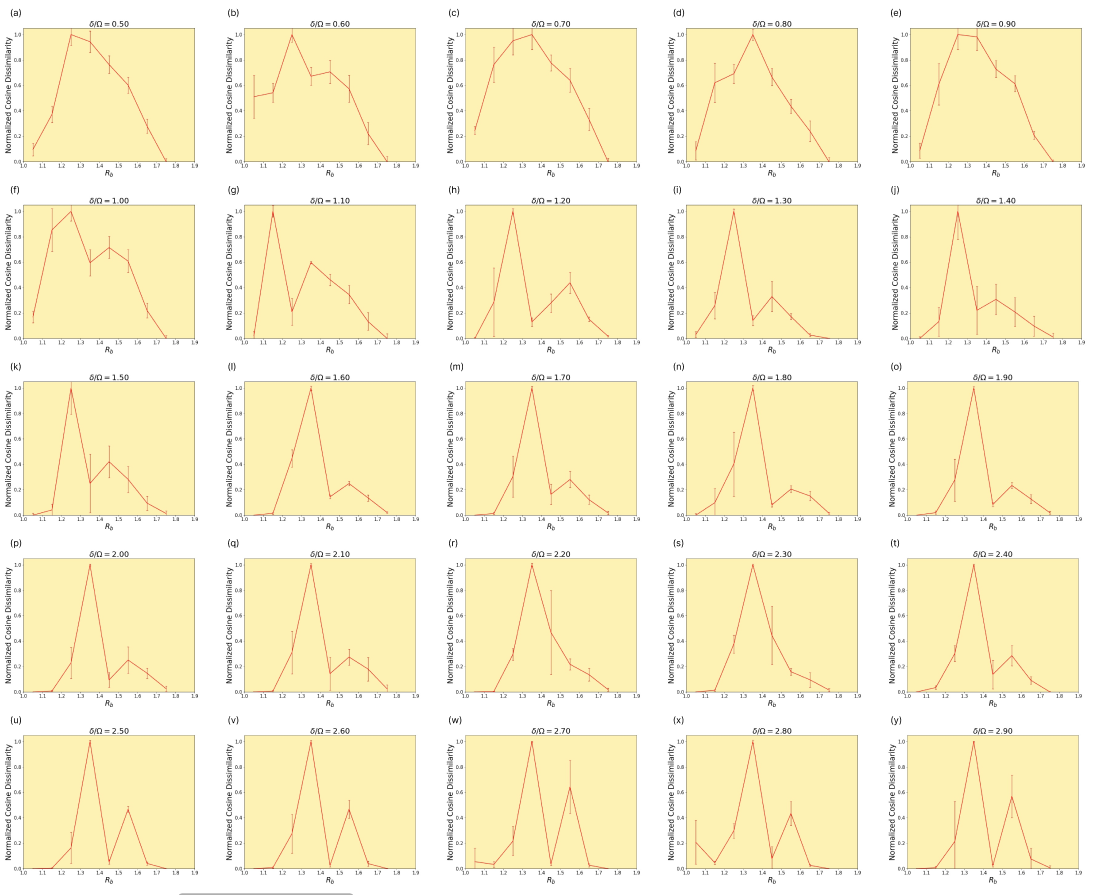}\\
\caption{We apply the SNN phase recognition method to the Rydberg array in the region $R_b \in \{1.0, \ldots, 1.8\}$ and $\delta \in \{0.5, \ldots, 2.9\}$. We scan along each vertical slice by choosing a fixed $\delta$, and traversing along all $R_b \in \{1.00, \ldots, 1.80\}$. We choose $k=1$ to perform adjacency comparisons within these scans. The dissimilarity peaks when configurations at $R_{b,i}$ and $R_{b,i+1}$ are of different phases.}
\label{fig:delta}
\end{center}
\end{figure}
\newpage

\section{Additional Adjacency Comparisons}
This section contains further examines the effect of the hyperparameter $k$ on the adjacency comparison.

\begin{figure}[htb!]
\begin{center}
\includegraphics[width=0.8\textwidth]{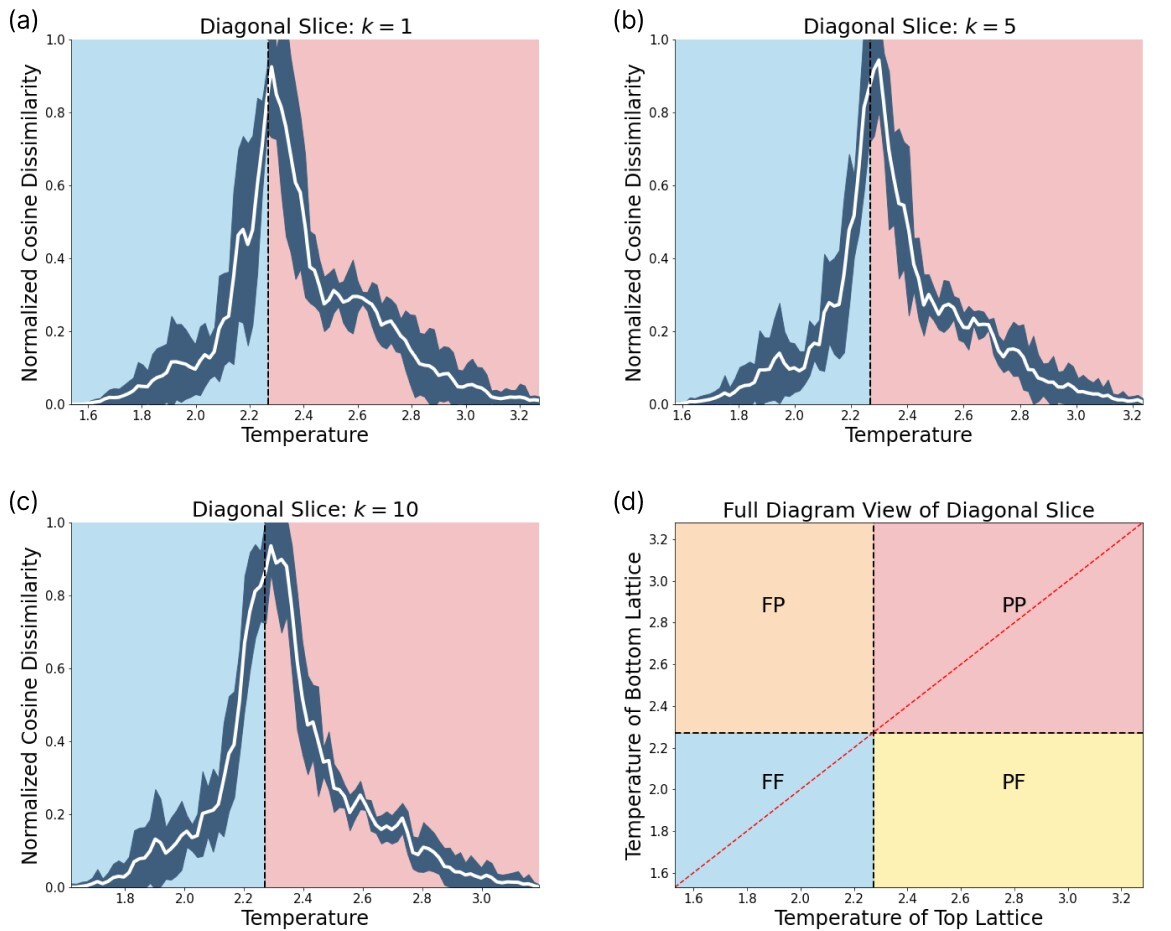}\\
\caption{We scan across the configurations of the stacked Ising Model belonging to the diagonal slice (d).  The adjacency comparison plots (a)-(c) depict the results of this scan using three different values of $k$. The diagonal line in (d) only crosses the phase boundaries once in the center (FF to PP), and we correctly observe a single corresponding peak in (a)-(c). Higher values of $k$ appear to smooth out noise. The trajectory of (a) is interrupted by a hump between $T=2.4...2.7$, which gradually lessens through to (c).}
\label{fig:diagonalDirect}
\end{center}
\end{figure}

\begin{figure}[htb!]
\begin{center}
\includegraphics[width=0.8\textwidth]{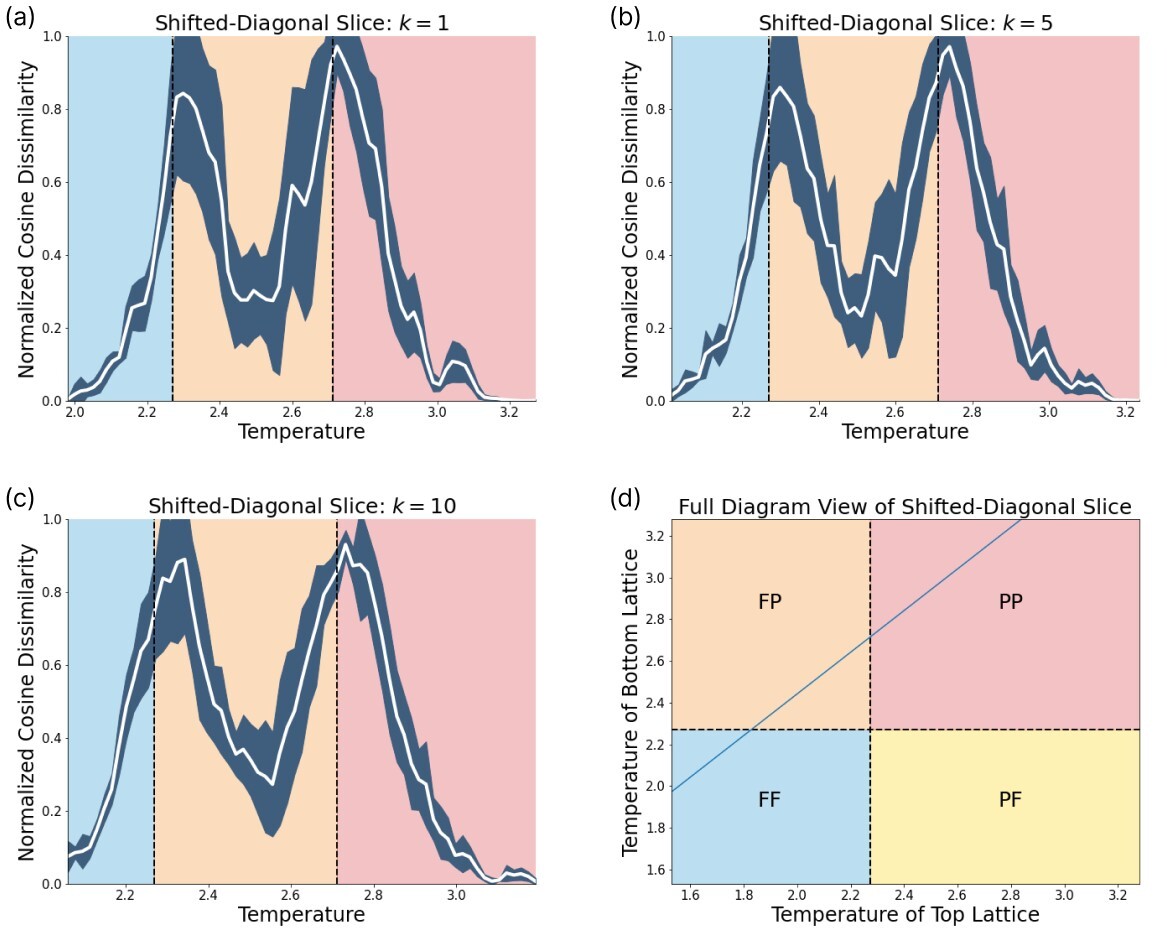}\\
\caption{We scan across the configurations of the stacked Ising Model belonging to the shifted diagonal slice (d). The adjacency comparison plots (a)-(c) depict the results of this scan using three different values of $k$. The diagonal line in (d) only crosses the phase boundaries once in the center (FF to PP), and we correctly observe a single corresponding peak in (a)-(c). Higher values of $k$ appear to smooth out noise. For instance, the FP phase of (a) is characterized by noise, which gradually lessens through to (c).}
\label{fig:shifted}
\end{center}
\end{figure}

\label{app:lt}

\newpage

\bibliographystyle{iopart-num}
\providecommand{\newblock}{}

%\bibliography{SiameseBib}

%%%%%%%%%%%%%%%%%%%%%%%%%%%%%%%%

\end{document}